\documentclass[superscriptaddress,amsmath,twocolumn,amssymb,prd,preprintnumbers,showpacs]{revtex4}
\usepackage{soul}
\usepackage{cancel}
\usepackage{amsmath,amssymb}
\usepackage{amsopn}
\usepackage{color}
\usepackage{float}
\usepackage[cmtip,arrow]{xy}
\usepackage{pb-diagram,pb-xy}
\usepackage{slashed}
\usepackage{graphicx}
\newcommand{\sla}[1]{\hbox{{$#1$}\llap{$/$}}}

\def \be {\begin{equation}}
\def \ee {\end{equation}}
\def \bea {\begin{eqnarray}}
\def \eea {\end{eqnarray}}

\def \dag {\dagger}

\def \mbp {\mathbf p}

\def \L {\lim_{V \rightarrow \infty}}
\begin{document}%........................................Beginning..................
\title{ Unitarity and Lee-Wick prescription at one loop level
 in the \\effective Myers-Pospelov electrodynamics: The $e^++e^-$ annihilation  }
\author{Carlos M. Reyes }
\email[Electronic mail: ]{creyes@ubiobio.cl  }
\affiliation{Departamento de Ciencias B{\'a}sicas, Universidad del B{\'i}o-B{\'i}o, Casilla 447, 
Chill\'an, Chile.}
\author{L. F. Urrutia}
\email[Electronic mail: ]{ urrutia@nucleares.unam.mx  }
\affiliation{Instituto de Ciencias Nucleares, Universidad Nacional Aut{\'o}noma de M{\'e}%
xico, 04510 M{\'e}xico D.F., M{\'e}xico. }
%\date{}
\begin{abstract}
We study perturbative unitarity in a Lorentz-symmetry-violating QED model with higher-order 
derivative operators in the light of the results of Lee and Wick to preserve unitarity in
indefinite metric theories. Specifically, we consider the fermionic sector of the 
Myers-Pospelov model, which includes dimension-five operators, coupled 
to standard photons. We canonically quantize the model, paying attention to its effective character, and show 
that its Hamiltonian is stable, emphasizing the exact stage at which  
the indefinite metric appears and decomposes into a 
positive-metric sector and a negative-metric sector.
Finally, we verify the optical theorem at the one-loop level in the 
annihilation channel of the forward-scattering process
$e^+(p_2, r) + e^-(p_1,s)$ by applying the Lee-Wick prescription in which 
the states associated with the negative metric are left out from the asymptotic Hilbert space, but nevertheless are considered in the loop integration via the propagator.
\end{abstract}
%\begin{keyword}
%Effective field theory, Lorentz violation, Higher derivative theory, Causality
%\end{keyword}
\maketitle
%-------------------------------------------------------------------------
\section{Introduction}
Gravitational effects of elementary particles are expected to
become significant at energy scales of the Planck mass $m_P\approx10^{19}$ GeV.
To describe the interplay between gravity and matter at 
these energies and to search for new physics, 
an effective approach has been actively exploited in the 
absence of a more fundamental theory.
A class of gravitationally induced effects which could be 
observable at standard-model energies
is the breakdown of Lorentz symmetry, which nevertheless 
is Planck-mass suppressed.
Many experiments have been designed to possibly detect 
such weak signals, and they range from precision laboratory
experiments to astrophysical observations. 

The standard-model extension (SME) is an effective framework 
that incorporates all possible Lorentz-invariance-violating terms for matter and gravity
in the Lagrangian. The breakdown of Lorentz symmetry 
originates from preferred directions
which are believed to arise from
expectation values of tensor fields in a more fundamental theory. 
A great number of the phenomenological searches for Lorentz symmetry violation
has been codified within the framework 
of the SME~\cite{SME,Table}. 
Originally, it was constructed to include
only renormalizable mass-dimension operators,---i.e, with dimension $d\leq 4$.
Recently, a generalization of the  
SME incorporating higher-order derivative operators has been proposed. 
Such a program has been successfully implemented in the 
photon sector~\cite{KM1}, the fermion sector~\cite{KM2}, and more recently in the linearized sector 
of gravity~\cite{KM3}. 

The pioneering work of Myers and Pospelov focuses on Lorentz invariance violation
with dimension-five operators coupled to a constant four-vector $n_\mu$ and having cubic 
dispersion relations in the lowest-order momentum expansion~\cite{MP}. The
Myers-Pospelov (MP) model has been studied to extract bounds upon its parameters from 
radiative corrections~\cite{Reyes,Manoel,rad2,Mariz}, cosmological 
observations~\cite{MP-Limits}, anisotropies ~\cite{anisotropies}, synchrotron radiation~\cite{syn} 
and also to analyze stability and causality~\cite{stability}. 
One can show that for a special choice of nonminimal SME coefficients, one arrives at the MP model.
Recently, an approach to introduce higher-order Lorentz symmetry violation, which lies beyond the
scope of the nonminimal SME with modified terms quadratic in the fields,
has been proposed with 
higher-order coupling terms~\cite{CouplingLIV}.

The interest in higher-order derivative operators in quantum field theory 
dates back to the work of Podolsky~\cite{pod}. He considered a higher-order
 electrodynamics to deal with infinities arising from the introduction
 of point charges.
 Some years later, Pais and Uhlenbeck realized
 that these higher-order derivative terms may lead to some problems with stability~\cite{PU}.
 The breakthrough in relation to stability came with the studies of Lee and Wick
 in the context of quantum field theories with an indefinite metric~\cite{LW}. 
 They give an important insight into the relation between the
possible loss of unitarity 
 and the interplay between statistics, stability and negative norm states.
Recently, the Lee-Wick ideas have been applied 
to solve 
 the hierarchy problem in the standard model~\cite{G}, to study the spectrum
of cosmological perturbations~\cite{cosmology} and to construct 
 a renormalization program in higher-derivative gravity~\cite{Modesto}. Also, higher-order derivative operators
have been included in quantum gravity approaches~\cite{Stelle,horava}, in
anisotropic regularization schemes~\cite{visser},
and in semiclassical gravity~\cite{S},
 and they arise in the study of the phenomenology 
of loop quantum gravity~\cite{loop} and in string theory~\cite{string}.

In 1969, T.~D.~Lee and G.~W.~Wick proposed a modified 
QED model with the advantage of being finite, but leading to an indefinite metric
in Hilbert space~\cite{LW}. They provide the main ideas towards
the construction of an indefinite metric quantum field theory with a unitary $S$ matrix.
The indefiniteness of the metric in the Lee-Wick quantum electrodynamics 
comes from a non-Hermitian Hamiltonian, which, however can be seen to arise from 
the presence of a higher-order derivative term as well~\cite{LWmodel}.
Several issues regarding stability and unitarity
were solved using what is now called
the Lee-Wick prescription. The analysis was extended by Cutkosky
using covariant perturbation 
theory based on Feynman diagrams~\cite{Cut}. 

The origin of the possible loss of unitarity
 in an indefinite metric theory can be found in the definition of the inner product.
To see this, consider two arbitrary states 
$\vert \phi \rangle=\sum_i   \phi_i \vert i \rangle   $ and $\vert \psi \rangle =\sum_j  
\psi_j \vert j \rangle$ expanded in a basis $\vert i \rangle$ with $i=1,2,3,\dots$
and $\phi_i,\psi_j$  
complex numbers. As 
 in usual quantum mechanics, the inner product between the two states is defined 
 by $\langle \phi \vert \psi\rangle=\phi^*_i \eta_{ij}\psi_j$,
where the metric $\eta=(\eta_{ij})$ is assumed to be a nonsingular Hermitian matrix
 and where the asterisk denotes complex conjugation.
Now, however, the generalization consists to allow for an
 indefinite metric, such that
 the diagonal terms of $\eta_{ij}=\langle i \vert  j\rangle$ can take negative values. 
 In this way, the metric $\eta$ in the Hilbert space is not positive definite,
and one may have states with a negative norm or ghosts in the theory.
The extended inner product induced by the 
indefinite metric $\eta$ in general leads to
a pseudo-unitary condition for the $S$ matrix, ---i.e.,
$S^{\dag}\eta S=\eta$. 
However, it was shown by Lee-Wick that by
removing the negative-metric particles from the asymptotic observable states of the 
 theory and defining a suitable choice for
the position of the poles 
in each Feynman diagram, the unitarity 
 of the $S$ matrix can be preserved~\cite{boulware,tonder}.
Importantly, one can show that in an indefinite metric theory,
 the algebra of creation and annihilation operators completely
determines the class of metrics $\eta$. 
For example, in a fermion system, once the Lagrangian is given, the equal-time 
anticommutation relations
lead to a unique metric representation. However, for bosons
using a redefinition of the vacuum state, one may change 
 from an indefinite metric representation to a positive definite metric representation, which however 
 may lead to instabilities.

Studies on perturbative unitarity in Lorentz-violating theories
have been carried out
in the renormalizable sector~\cite{unitarity1} and the nonrenormalizable sector~\cite{tree}.
At tree level, the conservation of energy plays a key role to verify
unitarity when higher-order derivative operators are present~\cite{tree}.
However, the preservation of unitarity is more involved when virtual 
particles are created, as in loop diagrams,
since then, one has to consider the discontinuities
of the poles associated with the particles with a negative metric as well.
In this work, using the ideas of Lee and Wick to deal with indefinite metric theories,
we extend 
some previous studies on unitarity to the one-loop level~\cite{unitarity}.

We consider the timelike MP model, where only the fermionic sector 
is modified with respect to standard QED, introducing higher-order time derivative 
operators yielding Lorentz symmetry violation. In Sec.~\ref{MD}, we provide 
the construction of the free fermionic field of the model, together with its basic 
properties: the calculation of the dispersion relations, the definition of the 
corresponding creation-annihilation operators, the verification of the anticommutation 
rules for the canonically conjugated field variables, the construction of  the 
Hamiltonian and the derivation of the free-field propagator using two different 
approaches---the vacuum expectation value of the time-ordered product 
$\psi(x)\,{\bar \psi}(y)$, and the integration over momentum space. We explicitly exhibit the  effective character  of the model by performing the quantization in a box, which leads to the exclusion of the excitations with imaginary frequencies. We 
follow closely the conventions of Ref.~\cite{MANDL}.

In Sec.~\ref{MPQED_U} we compute the imaginary part of the amplitude 
for the one-loop diagram in the annihilation channel arising from  the forward-scattering 
process $e^+(p_2, r) + e^-(p_1,s)$. The calculation is made using 
slightly modified  Feynman rules with respect to Ref.~\cite{MANDL}, which are 
stated at the beginning of this section. To this end, we calculate the 
amplitude ${\cal M}_F$ for the corresponding graph, and we also identify the 
integral $J_{\mu\nu}$ that produces the discontinuity in the amplitude ${\cal M}_F$, 
which yields the corresponding imaginary part. The contributions to such an integral are 
determined by the method of residues according to the 
appropriately defined  Lee-Wick contour, which is constructed by taking the same position of 
the poles which yields the correct answer when the propagator is calculated 
by integrating in momentum space. 

The discontinuities of ${\cal M}_F$ arising from 
$J_{\mu\nu}$ are subsequently obtained, yielding some 
unexpected cancellations, which nevertheless are crucial to prove the validity of 
the optical theorem in this case. Some details in the derivation of such discontinuities 
are given in the Appendix. In Sec.~\ref{VEROPT} we determine the amplitude 
${\cal M}_I$ for the process $e^+(p_2, r) + e^-(p_1,s) \rightarrow  e^+(k_2, {\bar r}) 
+ e^-(k_1,{\bar s})$ and calculate the sum over the momenta and spins of the  final 
states in $|{\cal M}_I|^2$  as required by the optical theorem. Unitarity 
is successfully verified by comparing this result with that obtained in the previous section 
for the imaginary part of ${\cal M}_F$. We close with Sec.~\ref{CONCL},
which contains our conclusions and final comments.
%------------------------------------------------------------------------

%............................................................................
\section{Model definitions} 
%............................................................................
\label{MD}
The modified QED Lagrangian we are interested in is obtained
via the minimal coupling substitution in the derivative terms  
of the fermionic sector in the Myers and Pospelov (MP) model,
\begin{eqnarray} 
\mathcal L=\bar \psi(i {\sla{D}}-m)\, \psi+g \bar \psi \sla{n} 
 (n\cdot D)^2\psi -\frac{1}{4}F_{\mu \nu}F^{\mu \nu}\,,
\label{MPED_LAG}
\end{eqnarray}
where $D_{\mu}= \partial _{\mu}-ieA_\mu$ and $n_\mu$ is a constant 
four-vector breaking the Lorentz symmetry and chosen in the timelike 
direction, such that  $n_\mu=(1,0,0,0)$. 
As usual, we will quantize this extended electrodynamics in the 
interaction picture, and we follow the conventions of Ref.~\cite{MANDL}.
We will work in the axial gauge $(n\cdot A)=0$, such that the interaction 
term is given by the standard one in QED. Also, the photon properties 
remain the same as in QED. On the contrary, the fermion sector will be 
drastically modified, and we start to study its properties in the free 
case. The corresponding equation of motion is 
\begin{eqnarray}  \label{eqm}
\left(i\sla{\partial} -m+g\gamma^0 \partial_0^2 \right)\psi(x)=0\,,
\label{FERM_EQ_MOT}
\end{eqnarray}
which includes higher-order time derivatives. Considering $\psi(x)=\int dp
 \,\psi(p)e^{-ip\cdot x}$, we obtain the eigenvalue equation
for the spinor field $\psi(p)$:
\begin{eqnarray} 
\left( \gamma^0(p_0-gp_0^2)+\gamma^ip_i -m \right)\psi(p)=0.
\label{EQ_MOT_MOM}
\end{eqnarray}
Taking the determinant of the above matrix
yields the dispersion relation
\begin{eqnarray}
(p_0-g p_0^2)^2-{\mathbf p}^2-m^2=0\,,
\label{DR}
\end{eqnarray}
whose solutions are
\begin{eqnarray} \label{field-energy}
\omega_1&=&\frac{ 1-\sqrt{1-4gE(\mathbf p)}}{2g}= \frac{1-N_1}{2g}\,,\nonumber  \\ W_1
&=&\frac{ 1+\sqrt{1-4gE(\mathbf p)  }}{2g}=  \frac{1+N_1}{2g}\,,
\label{ROOTS1} 
\end{eqnarray}
together with
\begin{eqnarray}
 \omega_2&=&\frac{ 1-\sqrt{1+4g E(\mathbf p)  }}{2g} =  \frac{1-N_2}{2g}\,, \nonumber  \\
W_2&=&\frac{ 1+\sqrt{1+4g E(\mathbf p)  }}{2g}= \frac{1+N_2}{2g}\,.
\label{ROOTS2} 
\end{eqnarray}
Here $E(\mathbf p)=\sqrt{{\mathbf p}^2+m^2}$,  $N_1=\sqrt{1-4gE(\mathbf p)}$ 
and $N_2=\sqrt{1+4gE(\mathbf p)}$. Let us observe that these functions are
 invariant under the change ${\mathbf p} \rightarrow - {\mathbf p}$. We identify the solutions
$\omega_1$ and $\omega_2$ with perturbations in $g$ of the usual ones
 $\pm E(\mathbf p)$, 
 while $W_1$ and $W_2$ correspond to  
 the contributions of new degrees of freedom coming from 
 higher-energy scales, which are nonperturbative in $g$. We emphasize that $\omega_2$ 
 is negative, so that the energy corresponding to this on-shell particle 
 is $|\omega_2|=-\omega_2$. The above eigenvalues satisfy the relation
\be
W_1+\omega_1= W_2+\omega_2\,.
\label{REL_FREQ} 
\ee 
The following additional identities follow from the definitions~\eqref{ROOTS1} and~\eqref{ROOTS2}:
\begin{eqnarray} 
E({\mathbf p}) &=&\omega_1- g\omega^2_1\,,  \nonumber  \\ E({\mathbf p}) &=&W_1-g W^2_1\,, 
\nonumber  \\ -E({\mathbf p}) &=&\omega_2- g\omega^2_2\,, \nonumber  \\  -E({\mathbf p})&=&W_2-g W^2_2\,,
\label{RELADD}
\end{eqnarray}
together with
\begin{eqnarray} 
E({\mathbf p})- g \omega_1^2 &=&\omega_1 N_1\,, \nonumber  
 \\ E({\mathbf p})- g W_1^2 &=&- W_1 N_1\,, \nonumber  
\\ E({\mathbf p})+g\omega_2^2&=&-\omega_2 N_2\,, \nonumber  \\ E({\mathbf p})
+g W_2^2&=&  W_2 N_2\,.
\label{RELADD1} 
\end{eqnarray}
From Eq.~\eqref{RELADD}, we observe that $(p_0-gp_0^2)=+E(\mathbf p)$ 
for $p_0=\omega_1$ or $p_0=W_1$, while $(p_0-gp_0^2)= - E(\mathbf p)$ 
for $p_0=\omega_2$ or $p_0=W_2$. In this way, the corresponding 
spinors satisfy the Dirac equations
\begin{eqnarray} 
\left( \gamma^0 E(\mathbf p)+\gamma^ip_i -m \right)u(\mathbf p)=0\,, 
\; {\rm for} \; p_0=\omega_1, W_1,
\label{DE1} 
\end{eqnarray}
and 
\begin{eqnarray} 
\left( \gamma^0 E(\mathbf p)+\gamma^ip_i +m \right)v(\mathbf p)=0\,, 
\; {\rm for} \; p_0=\omega_2, W_2,
\label{DE2} 
\end{eqnarray}
which correspond to the standard spinor solutions $u^r(\mathbf p)$ and $v^s(\mathbf p)$
labeled with the spin index $r,s$.
Our conventions for the completeness relations are
\begin{eqnarray} \label{compl1}
\sum _r u^r(\mathbf p) \bar u^r( \mathbf p)&=&\sla {p}+m\,, \nonumber \\
\sum _{r^{\prime}} v^{r^{\prime}}(\mathbf p) \bar v^{r^{\prime}}(\mathbf p)&=&\sla {p}-m\,,
\label{COMP_REL}
\end{eqnarray}
while for orthogonality we have 
\begin{eqnarray}
u^{s\dag}(\mathbf p) u^r(\mathbf p)&=&2E_p\delta^{sr}\,, \nonumber  \\ 
v^{s\dag}(\mathbf p) v^r(\mathbf p)&=&2E_p\delta^{sr}\,,\nonumber \\
u^{s\dag}(\mathbf p) v^r(-{\mathbf p})&=&0,\nonumber \\
 v^{s\dag}(\mathbf p) u^r(-{\mathbf p})&=&0.
\label{ORTHO} 
\end{eqnarray}
The above relations can be equivalently written as
\begin{eqnarray}
{\bar u}^s(\mathbf p) u^r(\mathbf p)&=&2m \delta^{sr}\,, \nonumber   \\
{\bar v}^s(\mathbf p) v^r(\mathbf p)&=& -2m \delta^{sr}\,, \nonumber   \\
{\bar u}^s(\mathbf p) v^r(\mathbf p)&=&0, \nonumber   \\
{\bar v}^s(\mathbf p) u^r(\mathbf p)&=&0.
\label{ORTHO1} 
\end{eqnarray}
Before proceeding to the quantization, let us recall the effective character of the timelike Myers-Pospelov
model, which is vividly illustrated by the existence of the momentum cutoff 
$|\mathbf{p}|_{\mathrm{max}}=\sqrt{M^{2}-m^{2}}$, where $M=1/4g$ is a very
high mass ($M >> m$), possible of the order of the Planck mass,  which suppresses the
Lorentz violating effects to be consistent with the already determined
experimental bounds. In fact, beyond $|\mathbf{p}|_{\mathrm{max}}$, the
frequencies $\omega _{1}$  and $W_{1}$ become imaginary thus introducing
stability problems together with runaway solutions which make the
quantization inconsistent.

In order to avoid such situation and to justify the effective
character of the theory, we  quantize the model in a cubic box of side $L$, under the following assumptions: first we impose
the  standard periodic boundary conditions upon the momenta, such  that
\begin{equation}
p_{i}=\frac{2\pi }{L}n_{i},\;\;\;i=x,y,z,\;\;n_{i}=0,\pm 1,\pm 2,\cdots,\pm N, \cdots\;
\label{MOMDIS}
\end{equation}
where the maximum value of each  $n_{i}$  is $N_{\rm max} \rightarrow \infty$. In this way the set of discrete functions $\exp (i\mathbf{%
p\cdot x})/\sqrt{V}$ is complete and orthonormal inside the box. The second step is to make $L \rightarrow \infty \,\,\, (V=L^3 \rightarrow \infty)$ according to the prescription
\be
L = 2\pi\sqrt{\frac{3}{M^2-m^2}}\, N_{\rm max}.
\label{PRESCRIP}
\ee
This guarantees that
\begin{eqnarray}
E({\mathbf p})={\mathbf p}^2+m^2&=&\left(\frac{2\pi}{L}\right)^2 (n^2_x+n^2_y+n^2_z)\\ &&+m^2\leq
\left(\frac{2\pi}{L}\right)^2 3 N^2_{\rm max}\nonumber \\&& \leq \frac{M^2-m^2}{3 N^2_{\rm max}}3 N^2_{\rm max} + m^2 =M^2, \nonumber 
\label{ELESSM}
\end{eqnarray}
for all values of ${\mathbf p}$, in such a way that $\omega_1$ and $W_1$ are always real inside the box. A confirmation that the scale $M$ is the physical way of realizing the mathematical infinity in the effective model comes from the susbtitution of the relation (\ref{PRESCRIP}) into Eq. (\ref{MOMDIS}) which yields 
\be
p_i\approx \frac{M}{\sqrt{3}}\frac{n_i}{N_{\rm max}}.
\ee 
This implies $(p_i)_{\rm max} \approx M/\sqrt{3}$. The analogous substitution in the standard box quantization would produce $(p_i)_{\rm max} \approx \infty/\infty$. 
In the following we have not carried out in mathematical detail the two assumptions described  above, but we have proceeded in an heuristic fashion by restricting ourselves to the effective model where we take for granted that $E({\mathbf p}) \leq M=1/4g$, together with the condition that the sum over the momenta runs over the entire ${\mathbf p}$ space.
The following expressions related to the box quantization will be required:
\begin{eqnarray}
%\begin{split}
 \int d^3x\, e^{i({\mathbf p}-{\mathbf p}^\prime)\cdot{\mathbf x}}=V\delta_{{\mathbf p},{\mathbf p}^\prime}, \nonumber \\  \sum_{{\mathbf p}}e^{i{\mathbf p}\cdot( {\mathbf x}-{\mathbf x}^\prime)}=V\, \delta^3({\mathbf x}-{\mathbf x}^\prime), \nonumber \\ 
\Phi(x)=\lim_{V \rightarrow \infty}  \int_{-\infty}^{+ \infty}\frac{dp_0}{2\pi}\sum_{{\mathbf p}}\frac{1}{\sqrt{V}}e^{-ip_0 x^0 +i {\mathbf p}\cdot {\mathbf x}} \, {\tilde \Phi}(p), \nonumber \\ 
\lim_{V \rightarrow \infty} \sum_{{\mathbf p}}\frac{1}{V}\, F({\mathbf p})=\int_{- \infty}^{+ \infty}\frac{d^3p}{(2\pi)^3} F({\mathbf p}). \label{CONVDISCR} 
%\end{split}
\end{eqnarray}
The first relation above is just the box normalization of the plane wave, while the second one
is the statement of completeness. The third relation defines the discrete Fourier transform in the box, together with the continuum one with respect to time. The last equation yields the passage to the continuum once all the calculations have been performed.

Having in mind the discretization previously defined we make
the following  expansion of the fields inside the box:
%\begin{widetext}
\begin{eqnarray}
\psi(x)&=& \lim_{V\rightarrow \infty} \sum_{\mbp,s}  \frac{1}{\sqrt{2VE(\mathbf p)}} \left(\frac{u^s(\mathbf p)}{\sqrt{N_1}} \left(  a^s_p
e^{-i\omega_1x_0} + c^{s}_p  e^{-iW_1x_0} \right)\nonumber  \right.  \\ &\times& \left.   e^{i\mathbf p\cdot \mathbf x}  
+   \frac{v^s(\mathbf p)}{\sqrt{N_2}} \left(  b^{s\dag}_p    
 e^{-i\omega_2x_0}  
+  d^{s}_p  e^{-iW_2x_0}  \right) e^{-i\mathbf p\cdot \mathbf x}  \right), \nonumber 
\\
\bar \psi(x)&=& \L \sum_{\mbp,s}  \frac{1}{\sqrt{2VE(\mathbf p)}}  
\left( \frac{\bar u^s(\mathbf p)}{\sqrt{N_1}}\left( a^{s\dag}_p  
   e^{i \omega_1 x_0} + c^{s\dag }_p e^{i W_1 x_0}\right) \nonumber   \right.  \\ &\times& \left.   e^{-i\mathbf p
   \cdot \mathbf x}+ \frac{\bar v^s(\mathbf p)}{\sqrt{N_2}}  \left( b^{s}_p 
  e^{i\omega_2 x_0} + d^{s\dag }_p  e^{iW_2 x_0}  \right)e^{i\mathbf p\cdot \mathbf x} 
  \right),
	\label{FEXP}
\end{eqnarray}
%\end{widetext}
where we have ${\bar \psi}(x)=\psi^\dagger \gamma^0$, because the all frequencies are real inside the box.
The commutation relations for  
particles and antiparticles (with frequencies $\omega_1$ and $\omega_2$, respectively) are taken to be the usual anticommutators
\begin{eqnarray}
\{a^{s}_p,a^{r\dag}_q  \}=\{b^{s}_p,b^{r\dag}_q  \}=\delta^{sr}\delta_{\mbp, {\mathbf q}}\,,
\label{OPCR1}
\end{eqnarray}
while for the new excitations (with frequencies $W_1$ and $W_2$) we impose
\begin{eqnarray}
\{c^{s}_p,c^{r\dag}_q  \}=\{d^{s}_p,d^{r\dag}_q  \}=-\delta^{sr}\delta_{\mbp, {\mathbf q}}.
\label{OPCR2}
\end{eqnarray}
The action of the annihilation operators  on the vacuum is defined by
\begin{eqnarray}
\label{VAC_ACT}
a^s_p |0\rangle =0\,, \quad b^{s}_p |0\rangle =0,\nonumber  \\ \quad  c^{s}_p |0\rangle =0\,, 
\quad   d^s_p |0\rangle =0.
\end{eqnarray}
The anticommutation relations in Eqs.~\eqref{OPCR1} and~\eqref{OPCR2} 
display the exact stage at which the indefinite metric decomposes into a positive-metric sector and
a negative-metric sector. Of course, due to the negative sign in Eq.~\eqref{OPCR2}, one is led to
identify the particles created
with the operators $c^{r\dag}_q$ and $d^{r\dag}_q$ as those with 
a negative-metric, while those created with $a^{r\dag}_q$ and $b^{r\dag}_q $
as particles with a positive-metric. We will show in 
 the Sec.~\ref{Canonical variables} that
 Eqs.~\eqref{OPCR1} and~\eqref{OPCR2} are necessary in order
 to fulfill the equal-time anticommutation relations 
 given in Eq.~\eqref{ETCR}.
 
The choice of vacuum in Eq.~\eqref{VAC_ACT} 
leads to the usual interpretation of the field
$\psi(x)$ annihilating fermions with positive energy $\omega_1$ and 
creating antifermions with positive energy $|\omega_2|$. 
In addition, the field $\psi(x)$ annihilates negative-metric
fermions with positive energy $W_1$ and negative-metric
 antifermions with positive energy $W_2$.
We will show in Section~\ref{stability} that 
this choice of vacuum leads to a  
Hamiltonian bounded from below.
%------------------------------------------------------------------------

%............................................................................
%..................................................................................
\subsection{Canonical variables}\label{Canonical variables}
%.................................................................................
Here we deal with the canonical quantization for the purely timelike MP Lagrangian
\begin{eqnarray} 
\mathcal L=\bar \psi(i {\sla{\partial}}-m)\, \psi+g  \psi^{\dag} \ddot \psi\,,
\label{MPL_TL}
\end{eqnarray}
which is hermitian, up to a total derivative, in virtue of the field definitions in Eq. (\ref{FEXP}).
The  canonically conjugated momenta to $\dot\psi$,
\begin{eqnarray}
\pi_{\dot \psi}=\frac{\partial \mathcal L}{\partial  \ddot \psi}\,, 
\end{eqnarray}
and to $\psi$,
\begin{eqnarray}
\pi_{\psi}=\frac{\partial \mathcal L}{\partial  \dot \psi}-\frac{\partial \pi_{\dot \psi}}{\partial  t}\,,
\label{CCMOM1} 
\end{eqnarray}
are given by
\begin{eqnarray}
\pi_{\psi}=i\psi^{\dag}-g \dot \psi^{\dag}\,, \nonumber \\
 \pi_{\dot \psi}=g \psi^{\dag}\,.
\label{CCMOM2}
\end{eqnarray}
Starting from the relations \eqref{FEXP}, \eqref{OPCR1}, and \eqref{OPCR2},
 and after taking the limit $V \rightarrow \infty$,
 we have verified the following equal-time commutation relations:
\begin{eqnarray}\label{comm1}
\{ \psi(x),\pi_{\psi}(y) \}&=&i\delta^3(\mathbf  x-\mathbf  y), \nonumber \\
\{ \dot \psi(x),\pi_{\dot \psi}(y) \}&=&i\delta^3(\mathbf  x-\mathbf  y)\,,
\label{ETCR}
\end{eqnarray}
with the remaining ones being zero.
 %.................................Hamiltonian.........................................
\subsection{Stability}\label{stability}
%..........................................................................
The Hamiltonian density is obtained from the Legendre transformation
${\mathcal H} = \pi_{\psi}\dot \psi+\pi_{\dot \psi} \ddot \psi-\mathcal L$, leading to
the  Hamiltonian 
\begin{eqnarray}
 H=\int d^3x \left(-g\dot 
 \psi^{\dag}(x) \dot \psi(x)+\bar \psi(x)(-i \gamma^i \partial_i+m)\psi(x)\right). 
 \label{HAM_MP}\nonumber \\
\end{eqnarray}
In a calculation analogous to the standard fermionic case, we have 
verified  that the Hamiltonian can be written as
 \begin{eqnarray}
 H= \sum_{\mbp, s} \left( \omega_1  
 a_p^{s\dag}   a_p^{s} +    
 \omega_2 b_p^{s} b_p^{s\dag}   - W_1  c_p^{s\dag}   c_p^{s}  
 - W_2     d_p^{s\dag}  d_p^{s} \right)\nonumber \\
\label{HAM}
 \end{eqnarray}
in terms of the creation-annihilation operators.
In obtaining Eq.~\eqref{HAM}, the identities
({\ref{RELADD1}) have been repeatedly used. Since $\omega_2 < 0$ and 
$\omega_1>0$, the above Eq.~\eqref{HAM} provides the usual interpretation 
for particle and antiparticle states: $a_p^{s \dagger}$ creates particles with 
momentum ${\mathbf p}$, spin component $s$ and energy 
$\omega_1({\mathbf p})$ and $b_p^{s \dagger}$ creates antiparticles 
with momentum ${\mathbf p}$, spin component $s$ and energy 
$|\omega_2 ({\mathbf p})|$, both of which are observable. A similar 
interpretation is given for the operators $c_p^{s\dag}$ and $d_p^{s \dagger}$ 
in terms of particles described by states with a negative metric.

 Introducing the number operators for particles 
in states with a positive metric,
$\hat N_{1p}= \sum_{s}  a_p^{s\dag}   a_p^{s}$, $\hat  N_{2p}= \sum_{s}  b_p^{s\dag}   b_p^{s}$,
and for particles 
in states with a negative metric,
$\hat  { \mathcal N}_{1p}= -\sum_{s}  c_p^{s\dag}   c_p^{s}$, 
$\hat  { \mathcal N}_{2p}=- \sum_{s}  d_p^{s\dag}   d_p^{s}$,
we can write
 \begin{eqnarray}
 H=\sum_{\mbp} \left( \omega_1
 \hat  N_{1p} -\omega_2  \hat  N_{2p}   + W_1   \hat{ \mathcal N}_{1p}
  +W_2   \hat  {\mathcal N}_{2p} \right)\,,
 \end{eqnarray}
 which is clearly bounded from below after dropping the usual infinite constant.
The Hamiltonian is stable and hermitian in the effective model.\\

%............................................................................
 %.................................Propagator........................................
\subsection{The propagator}
%...........................................................................................
Here we derive the fermion propagator $S_F(x-y)$ for the purely timelike MP model 
defined by the Lagrangian~\eqref{MPL_TL}. To verify that
its four-momentum representation (where one specifies the position 
of the poles $\omega_1, \omega_2, W_1, W_2$  in the complex 
$p_0$ plane and hence the contour integration $C_F$)  is correct,
 we first calculate the propagator according to 
 its definition in terms of the
vacuum expectation value 
$S_F(x-y)=\langle 0|T\left\{\psi(x) {\bar \psi}(y)\right\} | 0 \rangle$,
yielding
 \begin{eqnarray} 
S_F(x-y)&=&\theta(x_0-y_0)  \langle  0 |  \psi(x)  \bar \psi(y)  | 0\rangle - \theta(y_0-x_0)
\nonumber \\   &\times& \langle  0 |  \bar \psi(y) \psi(x) | 0\rangle\,.
\label{PROP_DEF}
\end{eqnarray}
Without loss of generality, we can set $y=0$. First, we consider $x_0>0$, 
and hence we need to calculate  $S_F^{(>)}(x)=  \langle  0 |  \psi(x)  \bar \psi(0)| 0\rangle $.
Using the expressions for the fields in Eq.~\eqref{FEXP}, we find
%\begin{widetext}
 \begin{eqnarray} 
S_F^{(>)}(x)&=& \L \langle  0 |  \left[ \sum_{\mbp, s} \frac{1}{\sqrt{2VE(\mathbf p)}} 
  \left( \frac{a^s_pu^s(p)}{\sqrt{N_1}} 
e^{-i\omega_1x_0+i\mathbf p\cdot \mathbf x}  \right. \right. \nonumber \\ &+& \left. \left. \frac{c^{s}_p u^s(p)  
 }{\sqrt{N_1}} e^{-iW_1x_0+i\mathbf p\cdot \mathbf x}  +\frac{d^{s}_p v^s(p)  
 }{\sqrt{N_2}} e^{-iW_2x_0-i\mathbf p\cdot \mathbf x}  \right) 
 \right.  \nonumber \\ 
 &\times& \left. \sum_{\mbp^\prime, r}  \frac{1}{\sqrt{2VE({\mathbf p^\prime})}}  
  \left( \frac{a^{r\dag}_{p^{\prime}}\bar  
u^r(p^{\prime})}{\sqrt{N_1^{\prime}}}   \right. \right. \nonumber \\ &+& \left. \left. \frac{c^{r\dag}_{p^{\prime}} \bar  u^r(p^{\prime})  
 }{\sqrt{N_1^{\prime}}} +\frac{d^{r\dag}_{p^{\prime}} \bar  v^r(p^{\prime})  
 }{\sqrt{N_2^{\prime}}}    \right)  \right]  | 0\rangle \,.
\label{PROP1}
\end{eqnarray}
%\end{widetext}
The properties in Eq.\eqref{VAC_ACT} allow us to rewrite  $ \langle
 0| X^{s}_p X^{r\dag}_q  |0\rangle =\langle 0|\{ X^{s}_p,  X^{r\dag}_q \}
  |0\rangle=\pm \delta^{r s} \delta_{{\mathbf p},{\mathbf q}}$, 
  where the plus sign is for $X=a,b$ and the minus sign for $X=c,d$. This yields
\begin{widetext}
\begin{eqnarray} 
S_F^{(>)}(x)= \L \sum_{\mbp,s} \frac{1}{2VE({\mathbf p})}   \left( \frac{u^s(p)  \bar  
u^s(p)}{N_1} \left(
e^{-i\omega_1x_0    +i\mathbf p\cdot \mathbf x}-e^{-iW_1x_0+i\mathbf p\cdot \mathbf x}
 \right) -\frac{ v^s(p)  \bar  v^s(p)  
 }{N_2}   e^{-iW_2x_0-i\mathbf p\cdot \mathbf x} \right)\,. 
\label{PROP2}
\end{eqnarray}
Using the completeness relation in Eq.~\eqref{COMP_REL}, we obtain
 \begin{eqnarray} 
S_F^{(>)}(x)&=& \L \sum_{\mbp} \frac{1}{2VE({\mathbf p})}   \left( \frac{\sla{p}+m  }{N_1}  \left( 
e^{-i\omega_1x_0+i\mathbf p\cdot \mathbf x}  - e^{-iW_1x_0+i\mathbf p\cdot \mathbf x}  \right)-
\frac{\sla{p}-m  }{N_2} e^{-iW_2x_0-i\mathbf p\cdot \mathbf x}
\right)\,,
\label{PROP3}
\end{eqnarray}
which can be rewritten as
\begin{eqnarray} 
S_F^{(>)}(x)&=&(i  \sla{\partial }   +m+g \gamma^0\partial_0^2   ) \L \sum_{\mbp} \frac{1}{2VE({\mathbf p})}  \left( \frac{e^{-i\omega_1x_0+i\mathbf p\cdot \mathbf x} }{N_1}
  -  \frac{e^{-iW_1x_0+i\mathbf p\cdot \mathbf x} }{N_1}  +\frac{ e^{-iW_2x_0-i\mathbf p\cdot \mathbf x} }{N_2} \right), 
\end{eqnarray}
%\end{widetext}
where the limit $L \rightarrow \infty$ can now be taken, yielding
\begin{eqnarray}
S_F^{(>)}(x)&=&(i  \sla{\partial }   +m+g \gamma^0\partial_0^2   )  \int \frac{d^3p}{2E({\mathbf p})(2\pi)^3}  \left( \frac{e^{-i\omega_1x_0+i\mathbf p\cdot \mathbf x} }{N_1}
  -  \frac{e^{-iW_1x_0+i\mathbf p\cdot \mathbf x} }{N_1}  +\frac{ e^{-iW_2x_0-i\mathbf p\cdot \mathbf x} }{N_2} \right), 
\label{PROP4}
\end{eqnarray}
\end{widetext}
In an analogous way, for $x_0<0$, we find
\begin{eqnarray} 
S_F^{(<)}(x)=- \L \sum_{\mbp,s} \frac{1}{2VE({\mathbf p})} \left( \frac{\sla{p}-m  }{N_2} 
e^{-i\omega_2x_0-i\mathbf p\cdot \mathbf x}  \right)\,,\nonumber \\
\label{PROP5}
\end{eqnarray}
which can be cast in the same form as Eq.~\eqref{PROP4}:
\begin{eqnarray} 
S_F^{(<)}(x)&=&(i  \sla{\partial}+m+g \gamma^0\partial_0^2) \int \frac{d^3p}{2E({\mathbf p})(2\pi)^3}  \frac{e^{-i\omega_2x_0-i\mathbf p\cdot \mathbf x}   }{N_2}\,.\nonumber \\
\label{PROP6}
\end{eqnarray}
Now we turn to the calculation of the propagator in momentum space.
From the equation of motion~\eqref{EQ_MOT_MOM}, 
the Feynman propagator is
\begin{eqnarray} 
S_F(x-y)=\L \int \frac{dp_0} {(2\pi)} \sum_{\mbp}\frac{1}{V}
 \frac{i(\sla{P}+m) }{P^2-m^2} e^{-ip\cdot( x-y)},\nonumber \\
\label{FPROP}
\end{eqnarray}
where $P^{\mu}=( p_0-gp_0^2,\mathbf p)$.
The propagator in momentum space is
\begin{eqnarray} 
S(p)=\frac{i (\sla{P}+m) }{P^2-m^2},
\label{FPROP_MOM}
\end{eqnarray} 
where one has to specify the position of the poles arising from the dispersion
 relations~\eqref{DR}, with solutions~\eqref{ROOTS1} and~\eqref{ROOTS2}, in 
 the complex $p_0$ plane. The standard $P^2-m^2 +i\epsilon $ prescription
  yields the following choice for the poles: 
$p_0=\omega_1-i\epsilon$, $p_0=W_1+i\epsilon$, $p_0=\omega_2+i\epsilon$, 
$p_0=W_2-i\epsilon$. Nevertheless, in order to reproduce the expressions 
(\ref{PROP4}) and (\ref{PROP6}) for the propagator  we have to choose the 
position for the corresponding poles as depicted in the following denominator of $S(p)$ 
\begin{widetext}
\begin{eqnarray} 
S(p)=\frac{i (\sla {P}+m)}{ g^2(p_0-(\omega_1-i\epsilon)) (p_0-(\omega_2+
i\epsilon))(p_0-(W_1-i\epsilon))(p_0-(W_2-i\epsilon))  }.
\label{NPP}
\end{eqnarray}
\end{widetext}

%...................................................Fig1......................
\begin{figure}
%\begin{figure}[H]
\centering
\includegraphics[width=0.55\textwidth]{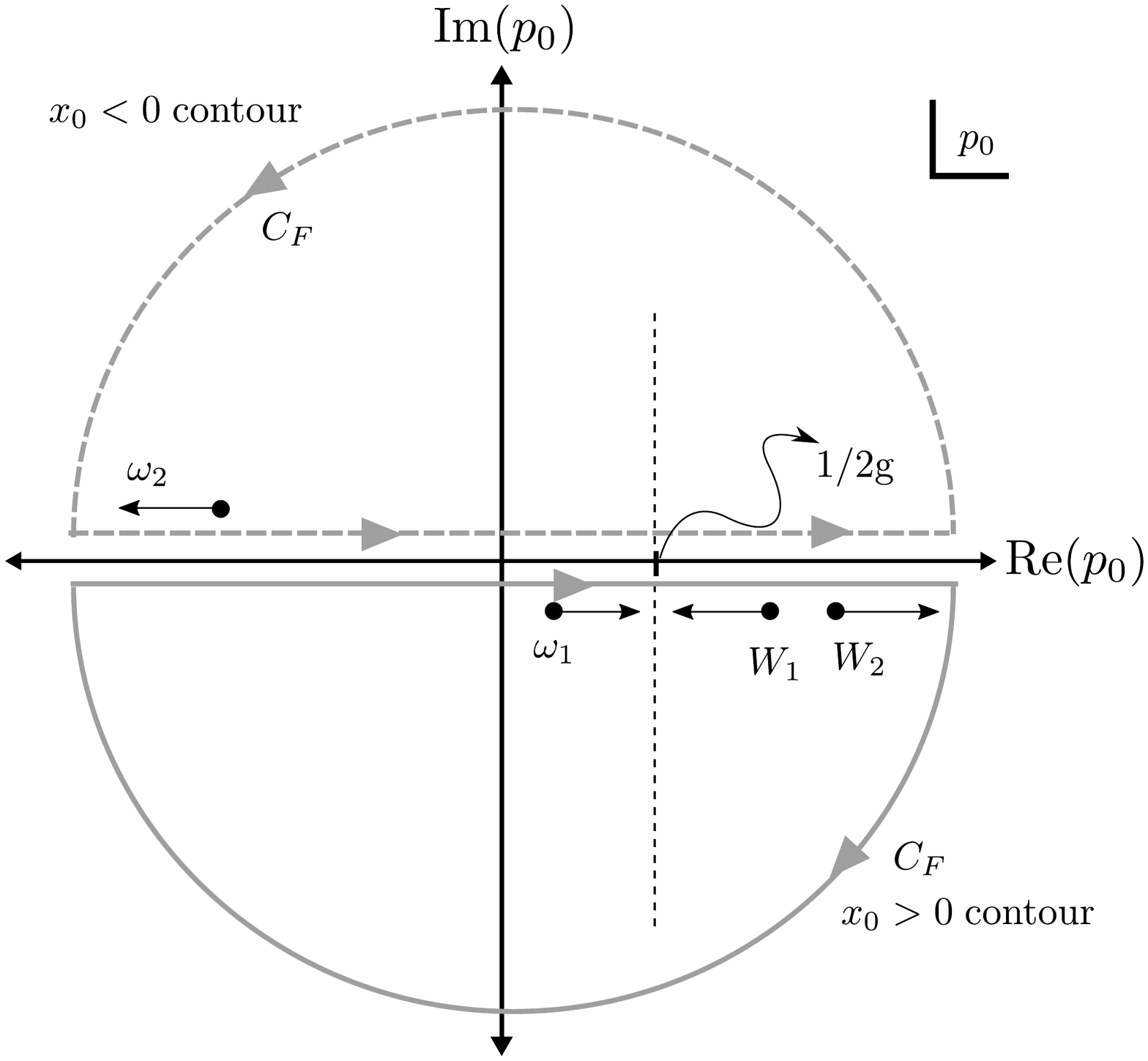}
\caption{\label{Fig1} The contour of integration $C_F$ which defines the Feynman propagator.
For $x_0>0$, it picks up the positive poles $\omega_1, W_1, W_2$, and for $x_0<0$, it
picks up the negative pole $\omega_2$. 
At the energy $E=1/4g$, the two 
poles $\omega_1$ and $W_1$ collide and move in opposite directions parallel to the imaginary axis as the energy increases. Nevertheless, our effective model is valid only for $E \leq 1/4g$, so that the case of  energies larger than $1/4g$ is not relevant  to our calculation. The poles 
$\omega_2$ and $W_2$ always stay in the real axis and move in opposite directions.}
\end{figure}
%............................................................................
That is to say, $\omega_1$, $W_1$ and $W_2$ are in the lower $p_0$ complex 
plane, while $\omega_2$ is in the upper $p_0$ complex plane. 
This choice is shown in Fig.~\ref{Fig1}, together with the motion of the poles as momentum
${\mathbf p}$ increases. The arrows indicate their trajectories in the $p_0$ 
plane according to Eqs.~\eqref{ROOTS1} and~\eqref{ROOTS2}. We see that 
both poles $\omega_1$ and $W_1$ collapse at the value $1/2g$ when 
$E({\mathbf p})=1/4g$ and move in opposite imaginary directions when 
$E({\mathbf p}) > 1/4g$. Since our effective model is defined only for $E \leq 1/4g=M$, such imaginary frequencies do not play any
 role in the calculation and are shown in Fig.(\ref{Fig1}) just for completeness. As $|{\mathbf p}|$ increases, the poles $\omega_2$ 
and $W_2$ move in real and opposite directions. In this way, the Feynman 
contour $C_F$ is defined as  the real axis with the poles located as shown in Fig.~\ref{Fig1}.
%\vspace{.5in}

Next, we derive the propagator using the expression (\ref{NPP}) together with the contour $C_F$
and show that we recover the expressions~\eqref{PROP4} and~\eqref{PROP6} obtained
 in the previous calculation by using its definition in terms of vacuum expectation values.
 To this end, let us set $y=0$ and 
 consider $x_0>0$. The factor $e^{i p_0\,x_0}$ in Eq.~\eqref{FPROP}
 indicates that we have to close our contour from below (${\rm Im} \, p_0 < 0$),
  thus enclosing the poles  $\omega_1$, $W_1$ and $W_2$. This yields
\begin{eqnarray} 
S^{(>)}_F(x)&=& \L \frac{i}{2\pi} \sum_{\mbp} \frac{1}{\sqrt{V}}  (-2\pi i)\left[  \left(\gamma^0E({\mathbf p})- 
 \mathbf \gamma \cdot \mathbf p+m \right) \right. \nonumber \\ &\times&\left. \left( \frac{e^{-i\omega_1
x_0+i\mathbf p\cdot \mathbf x}}{g^2(\omega_1-\omega_2)  (\omega_1-W_1)(\omega_1-W_2)    
 } \right. \right. \nonumber  \\
&+& \left. \left. \frac{e^{-iW_1
x_0+i\mathbf p\cdot \mathbf x}}{g^2(W_1-\omega_1)  (W_1-\omega_2)(W_1-W_2) } \right)
\right. \nonumber \\ &&\left. +(-\gamma^0E({\mathbf p})-  \mathbf \gamma \cdot \mathbf p+m )
\right.  \nonumber\\ &\times& \left. \frac{e^{-iW_2
x_0-i\mathbf p\cdot \mathbf x}}{g^2(W_2-\omega_1)  (W_2-\omega_2)(W_2-W_1) }  
\right]  \,. 
\label{INTBELOW}
\end{eqnarray}
Using  the identities
\begin{eqnarray} \label{id1}
g^2(\omega_1-\omega_2)  (\omega_1-W_1)(\omega_1-W_2)&=&2EN_1\,,
\label{ID1}
\nonumber \\ \label{id2}
g^2(W_1-\omega_1)  (W_1-\omega_2)(W_1-W_2)&=&-2EN_1\,, 
\label{ID2}  
\nonumber \\ \label{id4}
g^2(W_2-\omega_1)  (W_2-\omega_2)(W_2-W_1) &=&2EN_2 \,,
\label{ID4}
\end{eqnarray}
we have
\begin{eqnarray} 
S^{(>)}_F(x)&=&\L \sum_{\mbp} \frac{1}{V}    \Big( (\gamma^0E({\mathbf p})-
  \mathbf \gamma \cdot \mathbf p+m ) \\ &\times&   \left( \frac{e^{-i\omega_1
x_0+i\mathbf p\cdot \mathbf x}}{2EN_1    }- \frac{e^{-iW_1
x_0+i\mathbf p\cdot \mathbf x}}{2EN_1  }\right) \nonumber \\ && +(-\gamma^0E({\mathbf p})+
 \mathbf \gamma \cdot \mathbf p+m ) \frac{e^{-iW_2
x_0-i\mathbf p\cdot \mathbf x}}{2EN_2  }\Big)\,, \nonumber 
\label{INTBELOW1} 
\end{eqnarray}
where we have changed ${\mathbf p}\to - {\mathbf p}$ in the last term.
After taking the limit $V \rightarrow \infty$, the above expression reduces to
\begin{eqnarray} \label{prop1}
S^{(>)}_F(x)&=&(i  \sla{\partial }   +m+g \gamma^0\partial_0^2   )  \int \frac{d^3p }{2E
(2\pi)^3}  \Big(  \frac{e^{-i\omega_1
x_0+i\mathbf p\cdot  \mathbf x}}{N_1}  \nonumber\\ && - \frac{e^{-iW_1
x_0+i\mathbf p\cdot  \mathbf x}}{N_1}+\frac{e^{-iW_1
x_0-i\mathbf p\cdot  \mathbf x} }{N_2} \Big)\,.
\label{INTBELOW2}
\end{eqnarray}
The above expression reproduces the form of the propagator obtained in Eq.~\eqref{PROP4}.
When $x_0<0$, the $p_0$ integration is made by closing the contour from above, and we obtain
\begin{eqnarray} 
S^{(<)}_F( x)&=& \frac{i}{2\pi} \L  \sum_{\mbp} \frac{1}{V}   (2\pi i)\Big(  (-\gamma^0E({\mathbf p})
-  \mathbf \gamma \cdot \mathbf p+m )  \nonumber\\ &\times& \frac{e^{-i\omega_2
x_0+i\mathbf p\cdot \mathbf x}}{g^2(\omega_2-\omega_1)  (\omega_2-W_1)(\omega_2-W_2) 
    }\Big)\,.
\label{INTABOVE}
\end{eqnarray}
Using now  the relation
\begin{eqnarray} \label{id3}
g^2(\omega_2-\omega_1)  (\omega_2-W_1)(\omega_2-W_2) =-2EN_2 \,,
\label{ID3}
\end{eqnarray}
we finally arrive at
\begin{eqnarray} 
S^{(<)}_F(x)=(i  \sla{\partial} + m + g \gamma^0\partial_0^2) \int \frac{d^3 p}{2E(2\pi)^3} 
 \frac{e^{-i\omega_2x_0-i\mathbf p\cdot \mathbf x} }{N_2}\,,\nonumber  \\
\end{eqnarray}
which reproduces Eq.~\eqref{PROP6}. In this way, we have proved that the 
prescription~\eqref{NPP} for the position of the poles  in the complex $p_0$ 
plane yields the correct fermionic propagator, according to the definition in
Eq.~\eqref{PROP_DEF}. 

%............................................................................
%........Unitarity in the Myers-Pospelov Electrodynamics: one loop level.................
\section{Unitarity in the Myers-Pospelov Electrodynamics: one-loop level}
\label{MPQED_U}
%............................................................................................................................
Our goal is to verify the optical theorem for the simple diagram shown in 
Fig.~\ref{Fig2}. In the conventions of Ref.~\cite{MANDL} the $S$ matrix is defined as
\begin{eqnarray} 
S\equiv  1+ iT&=&1+\left( 2\pi \right) ^{4}\delta ^{4}(P_{i}-P_{f})\prod_{i}\left( \frac{%
1}{2VE_{i}}\right) ^{1/2}    \nonumber  \\ &\times& \prod_{f}\left( \frac{1}{2VE_{f}}\right)
^{1/2}(i\mathcal{M}). \label{SM}
\end{eqnarray}
The mass fermions factors $\prod_{F}(2m_{F})^{1/2}$ appearing in the corresponding expression of Ref.~\cite{MANDL} have been incorporated in our definitions of the fermion wave functions $u$'s and $v$'s. Then we have the relation $u^s({\mathbf p})=\sqrt{2m}\,{U}^s({\mathbf p})$ and correspondingly for $v^s({\mathbf p})$, which transforms the spinor completeness and orthogonality relations for ${U}^s({\mathbf p})$ in Ref.~\cite{MANDL} into ours, given in Eqs. (\ref{COMP_REL}) and (\ref{ORTHO}).  We take the same definition and normalization of the one particle 
state as in Ref.~\cite{MANDL}, so
\begin{eqnarray} 
|e^-, {\mathbf p}, s \rangle&=&a_p^{s\dagger}| 0\rangle\,, \nonumber \\ 
|e^+, {\mathbf q}, r \rangle&=&b_q^{r\dagger} | 0\rangle \,, 
\label{OPS}
\end{eqnarray}
together with 
\be
\langle {\mathbf p}, s| {\mathbf q}, r  \rangle =\delta^{sr} \delta_{{\mathbf p},{\mathbf q}}\,.
\label{NORM}
\ee 
Then, the completeness relation for one particle states is 
\be
\sum_{{\mathbf p}, s}\,|{\mathbf p}, s\rangle \langle {\mathbf p}, s |=1.
\ee 
Following standard steps and taking the limit $V\rightarrow \infty$ according to the last expression in Eq.(\ref{CONVDISCR}), the required form of the optical theorem reads
\begin{eqnarray} 
&&2{\rm{Im}}{\cal M}(A \rightarrow A)=\sum_n \left(\prod_i^n \int 
 \frac{d^3 k_i}{(2 \pi)^3
} \frac{1}{2E({\mathbf k}_i)}\right) \nonumber \\ 
&\times& \sum_{{\bar s}_i}\, \left|{\cal M}(A \rightarrow B(\textbf{k}_1, 
{\bar s}_1, \dots, \textbf{k}_n, 
{\bar s}_n ))  \right|^2 \nonumber \\ &\times& (2\pi)^4   \delta^4 \left(p_A-\sum_i k_i\right)\,.
\label{OPTTHEO}
\end{eqnarray}
Here $A$ denotes the initial and final processes associated with the amplitude
 ${\cal M}(A \rightarrow A) \equiv {\cal M}_F$, each process carrying total momentum 
 $p_A$. The term in the right hand side  ${\cal M}(A \rightarrow B(\textbf{k}_1, 
{\bar s}_1, \dots, \textbf{k}_n, 
{\bar s}_n )) \equiv {\cal M}_I$ denotes the scattering 
 amplitude for the process $A$ going to the final states of $n$ particles
 described 
 by $B(\textbf{k}_1, 
{\bar s}_1, \dots, \textbf{k}_n, 
{\bar s}_n ))$. In our case, the process 
 $A$ corresponds to
 $e^-(p_1, s)+ e^+ (p_2, r) $, with $p_A=p_1+p_2$ and the  process $B$ is 
$ e^-(k_1, {\bar s})+ e^+ (k_2, {\bar r})$, as shown in Fig.~\ref{Fig2}.
%------------------------------------------------------------------------
%............................................................................
%...................................................Fig2......................
\begin{figure}
\centering
\includegraphics[width=0.5\textwidth]{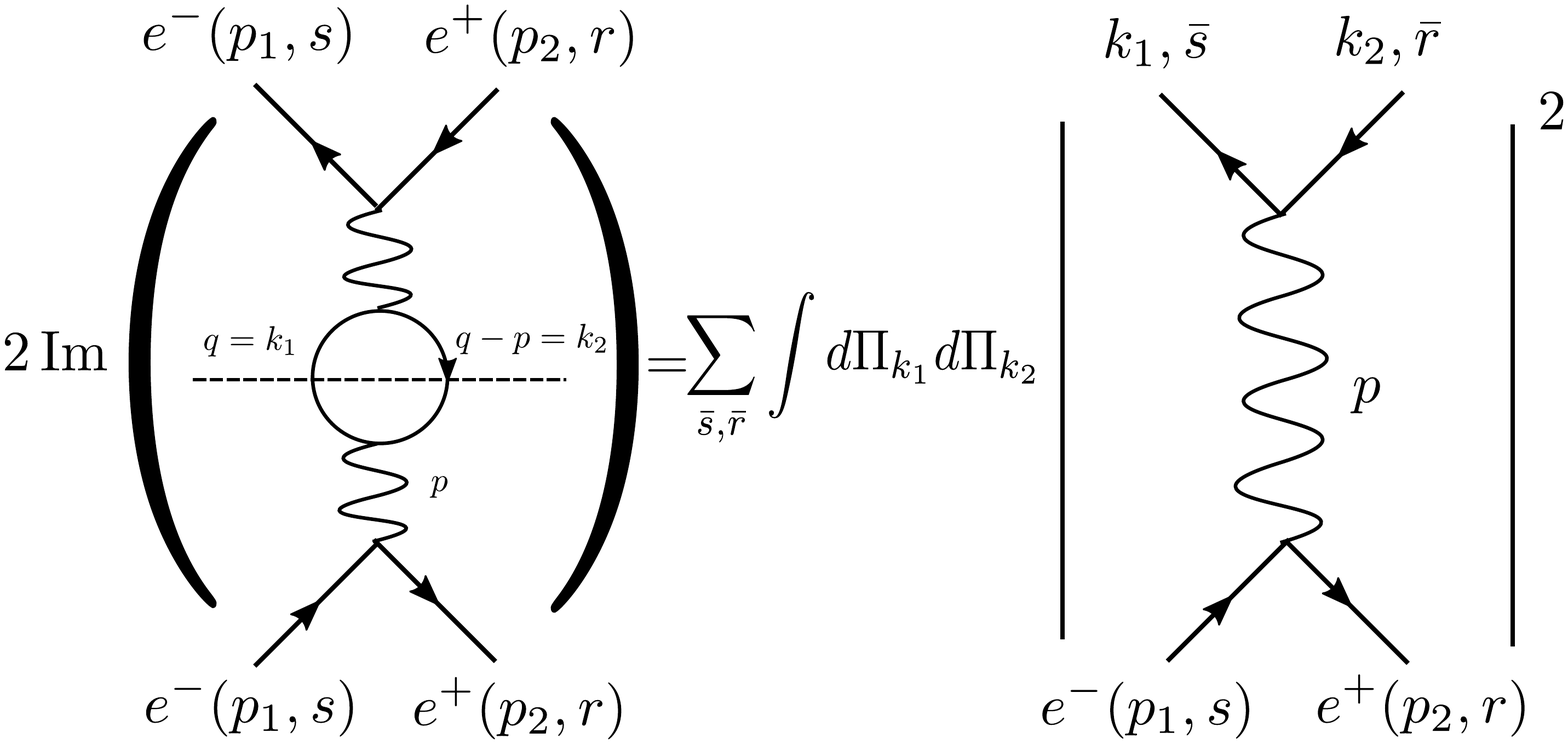}
\caption{\label{Fig2} The optical theorem, showing the forward-scattering process 
$e^-(p_1, s)+ e^+ (p_2, r)$ on the left-hand side and
the sum over intermediate states 
$e^-(k_1, {\bar s})+ e^+ (k_2, {\bar r})$, conserving total 
energy $p_{A}$, on the-right hand side.  The phase-space measure 
is defined by $
\int d\Pi_{k_1} d\Pi_{k_2}= \int 
\left( \frac{d^3 k_1}{(2 \pi)^3
} \frac{1}{2E({\mathbf k}_1)}\right)  \left( \frac{d^3 k_2}{(2 \pi)^3
} \frac{1}{2E({\mathbf k}_2)}\right) (2\pi)^4 
\delta^4\left( p_A-( k_1 +  k_2 \right) $. 
}
\end{figure}
%......................................................................................

We have adopted the Lee-Wick prescription, where the intermediate 
states on the right-hand side of Eq.~\eqref{OPTTHEO} include only the
positive-metric states---i.e.,  $e^-$ and $e^+$ in our case---while leaving
out the negative-metric states~\cite{LW}. Nevertheless, 
the latter states may contribute to the imaginary part of the left-hand side
via the propagator in the loop integral, and this mismatching 
could be a cause for the failure of unitarity. Still, as we show in the 
following, a proper choice of the so-called Lee-Wick integration
contour will restore unitarity at the level of the optical 
theorem~\eqref{OPTTHEO}.  

With the conventions indicated above, the required Feynman rules to calculate the above
 amplitudes are the same as those for QED in Ref.~\cite{MANDL}, 
 except for the following changes: (i) The photon propagator is 
\begin{equation}
D_{\mu\nu }(p)=\frac{-i}{p^2 +i\epsilon} \left[ \eta_{\mu\nu}-
\frac{p_\mu n_\nu +n_\mu p_\nu }{(n\cdot p)} +p_\mu p_\nu \frac{n^2}{(n \cdot p)^2}\right]\,,
\label{FOT_PROP}
\end{equation} 
which is given in the homogeneous temporal axial gauge. 
Nevertheless, in our case this propagator joins two conserved currents,
 in such a way that only the term proportional to $\eta_{\mu\nu}$ contributes. 
 Also, we know that the Fadeev-Popov ghosts required in the case of the 
 axial gauges decouple, so that there is no contribution coming from them. 
(ii) The fermion propagator is 
\begin{equation}
S(q)=\frac{i(\gamma_0 f(q_0)+ \gamma_i q^i +m )}{g^2(q_0-\omega_1)
(q_0-\omega_2)(q_0-W_1)(q_0-W_2)}\,,
\label{FERM_PROP1}
\end{equation}
with the notation 
\begin{equation}
f(q_0)= q_0 - g q_0^2\,,
\end{equation}
in agreement with Eq.~\eqref{FPROP_MOM}, where the position of the poles
has already been specified in Eq.~\eqref{NPP}.
Here we have contributions
from the negative-metric states via the poles $W_1$ and $W_2$. (iii) The last change 
in the Feynman rules comes from the fermionic external lines, where
we have to introduce the following replacements: 
\be
u^s(\mathbf p) \rightarrow \frac{u^s(\mathbf p)}{\sqrt{N_1(\mathbf p)}}, \qquad
v^s(\mathbf p) \rightarrow \frac{v^s(\mathbf p)}{\sqrt{N_2(\mathbf p)}}\,,
\label{NEWUV}
\ee
and analogously for ${\bar u}^s(\mathbf p)$ and ${\bar v}^s(\mathbf p)$. 
As established previously, the spinors $u^s(\mathbf p)$ and $v^s(\mathbf p)$ 
are the same as in standard QED. The first relation in (\ref{NEWUV}) can be 
directly seen from the action
\begin{eqnarray} 
\psi ^{+}(x)|e^{-},\mathbf{q},r\rangle& =&\sum_{s,\mathbf{p}}\frac{1}{\sqrt{2 V E(\mathbf{p})}}\frac{u_{s}(\mathbf{p})}{\sqrt{N_{1}(\mathbf{p})}}%
\nonumber \\  &\times& e^{-ip\cdot x} \, a^{s}{}_{p}a^{+r}{}_{q}|0\rangle \,,
\label{CHECK_NEWU}
\end{eqnarray}}
required in the Wick expansion of the interaction Hamiltonian. The property
\begin{equation}
a^{s}{}_{p}a^{+r}{}_{q}|0\rangle =\left\{ a^{s}{}_{p},\;a^{+r}{}_{q}\right\}
|0\rangle =\delta ^{rs}\delta_{\mathbf{p}, \mathbf{q}  }\,,
\label{ACREL}
\end{equation}
yields
\begin{equation}
\psi ^{+}(x)|e^{-},\mathbf{q},r\rangle =\left( \frac{1}{\sqrt{2 V E(\mathbf{q})}}\right)\left(\frac{u_{r}(\mathbf{q})}{\sqrt{N_{1}(\mathbf{q})}
} \right) e^{-i\omega _{1}(\mathbf{q})x_{0}}e^{i\mathbf{q\cdot x}}\,.
\label{EXT_FERM}
\end{equation}
The first parenthesis in the right hand side of the above equation has been factored out in the expression (\ref{SM}) for the $S$ matrix, and 
the resulting exponential contributes to the total energy-momentum 
conservation, with the 
physical energy $\omega _{1}(\mathbf{q})$ arising from the dispersion 
relations. Then, only the term $u_{r}(\mathbf{q})/\sqrt{N_{1}(\mathbf{q})}$ 
contributes to the amplitude $(i{\cal M})$.  A similar result  can be obtained for the remaining spinors, 
thus validating the replacements shown in Eq.~\eqref{NEWUV}.

The amplitude ${\cal M}_F$ for the graph shown in the 
left-hand side of Fig.~\ref{Fig2} is 
\begin{eqnarray} 
i{\mathcal M}_F&=&(-1)   \frac{1}{N_1({\mathbf p}_1) N_2({\mathbf p}_2)}
\bar{v}^r(p_{2})(-ie\gamma ^{\mu
})u^s(p_{1}) \nonumber \\ && D_{\mu\nu}(p)
\int \frac{d^{4}q}{(2\pi )^{4}}\; {\rm Tr} \Big[  (-ie  \gamma ^{\rho })
S (q-p)  (-ie \gamma ^{\nu })
S(q)  \Big]  \nonumber \\
&&\times  D_{\rho \sigma }(p) \,
\bar{u}^s(p_{1})(-ie\gamma ^{\sigma})
v^r(p_{2})\,,
\label{AMPLITUDE}
\end{eqnarray} 
where the minus sign comes from the fermion loop, and $p^\mu=p_1^\mu + p_2^\mu$. 
Let us define the currents
\begin{eqnarray}
J_{1}^{\mu }(p_1,p_2)&=&\frac{1}{\sqrt{N_1({\mathbf p}_1) N_2({\mathbf p}_2)}}
\bar{v}^r (p_{2})\gamma ^{\mu}u^s(p_{1})\,, \nonumber \\
J_{2}^{\mu }(p_1,p_2)&=& \frac{1}{\sqrt{N_1({\mathbf p}_1) N_2({\mathbf p}_2)}}
\bar{u}^s(p_{1})\gamma ^{\mu
}v^r(p_{2}) \nonumber \\ &=&\left[ J_{1}^{\mu }(p_1,p_2)\right] ^{\ast }\,.
\label{CURRENTS}
\end{eqnarray}
Due to current conservation at the ingoing and outgoing  vertices,  
$p_\mu J_{1}^{\mu }=0$ and $p_\mu J_{2}^{\mu }=0$, the only 
contribution from the photon propagator to the amplitude ${\cal M}_F$
 arises from the term containing $\eta_{\mu\nu}$. 

In the center-of-mass frame ($\mathbf p=0$) and using 
Eq.~\eqref{FERM_PROP1}, we can write
%\begin{widetext}
\begin{eqnarray}
{\cal M}_F&=&-\frac{ e^4}{p^4}J_1^\nu J_2^\mu 
\int \frac{d^{3}q}{(2\pi )^{4}} \int_{C_{LW}} dq_0 (-i) \;  \nonumber \\ &\times& \text{Tr}\left[ \gamma _{\mu }
\frac{(\gamma^0 {  f(q_0-p_0)}+ \gamma^i {  q_i}    +m) }{D_{q-p}} \gamma _{\nu } \right. \nonumber \\  &\times& \left. \frac{
(\gamma^0{ f(q_0)}+ \gamma^iq_i +m)}{D_{q}} \right] \,,
\label{AMP2}
\end{eqnarray}
%\end{widetext}
where 
\begin{eqnarray}
 D_{q} &=& g^2(q_0-\omega_1)(q_0-\omega_2)(q_0-W_1)\nonumber \\
 &\times&(q_0-W_2)\,, \\
  D_{q-p} &=& g^2(q_0-\widetilde \omega_1)(q_0-\widetilde \omega_2)
  (q_0-\widetilde W_1)\nonumber \\
 &\times&(q_0-\widetilde W_2)\,,
\label{INTI}
\end{eqnarray}
with
\begin{eqnarray} 
\widetilde \omega_1&=&p_0+\omega_1\,, \nonumber  \\   \widetilde \omega_2&=&p_0+\omega_2 \,,
\nonumber  \\ \widetilde W_1&=&p_0+W_1\,,\nonumber  \\ \widetilde W_2&=&p_0+W_2\,.
\end{eqnarray}
It is convenient to define
\begin{eqnarray}
T_{\mu\nu}(p_0, q_0,{\mathbf q})&\equiv& \text{Tr}\left[ \gamma _{\mu }
(\gamma^0 {  f(q_0-p_0)}+ \gamma^i {  q_i}    +m) \gamma _{\nu } \right. \nonumber \\
&\times& \left.
(\gamma^0{ f(q_0)}+ \gamma^iq_i +m) \right],
\label{DEFTMUNU}
\end{eqnarray}
and
\begin{eqnarray}
I(p_0, q_0,{\mathbf q})&\equiv&\frac{-i}{D_qD_{q-p}}\,,
\end{eqnarray}
together with
\be
J_{\mu\nu}(p_0, {\mathbf q})=\int_{C_{LW}} dq_0 T_{\mu\nu}(p_0, q_0,{\mathbf q})
 I(p_0, q_0,{\mathbf q}).
\label{DEFJMUNU}
\ee
We recall that the poles $\omega_1$ and $W_1$ and $W_2$ are in the lower complex 
$p_0$ plane, while $\omega_2$ is in the upper complex $p_0$ plane.

\begin{figure}[H]
\centering
\includegraphics[width=0.55\textwidth]{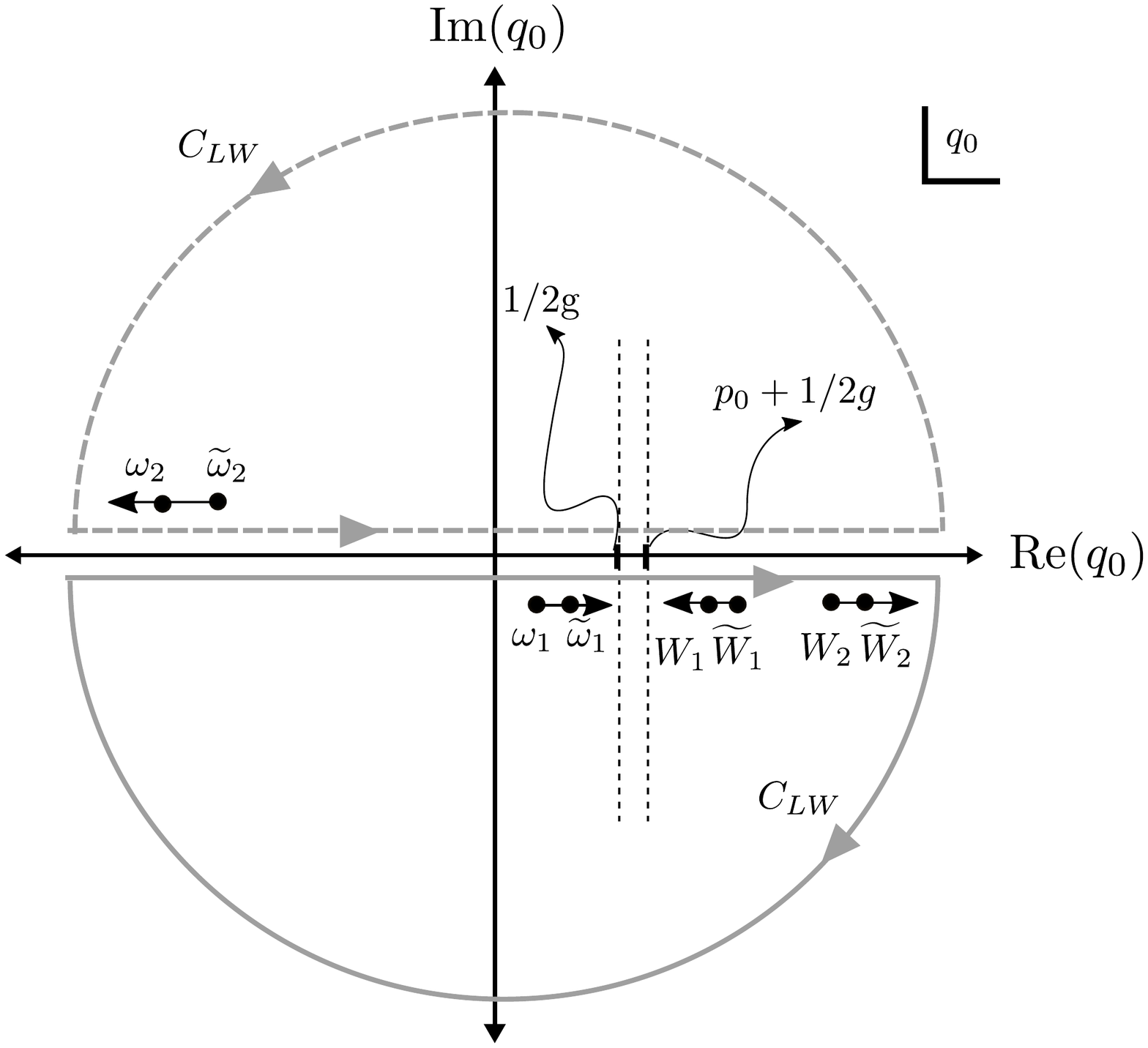}
\caption{\label{Fig3} The Lee-Wick contour $C_{LW}$ used to compute 
the imaginary part of the forward-scattering process ${\mathcal M}_F$. Also
shown is the trajectory 
of each pole as the energy increases. Only the case where  $E \leq 1/4g=M$ is relevant for the purposes of our effective model defined inside the box of volume $V$.}
\end{figure}

We define the corresponding Lee-Wick contour $C_{LW}$, shown in Fig.~\ref{Fig3}, 
such that the poles 
$\omega_1$, $W_1$, $W_2$,  $\widetilde \omega_1$, $\widetilde W_1$
and $\widetilde W_2$ are in the lower sector, while the poles $\omega_2$ and $\widetilde \omega_2$ are in the upper sector. Then we have two ways of calculating the
integral $J_{\mu\nu}(p_0,{\mathbf q})$ by closing the Lee-Wick contour in  the
upper or the lower complex $p_0$ plane. 

First, closing the contour downward yields 
\begin{eqnarray}
J_{\mu\nu}(p_0,{\mathbf q})&=&(-2\pi i)\sum_{z} T_{\mu\nu}(q_0, p_0, {\mathbf q})  ]_{
q_0=z} \nonumber  \\ &\times& [\text{Res} \, I(q_0, p_0, {\mathbf q})]_{q_0=z}\equiv  (-2\pi i)\nonumber 
\\ &\times& \sum_z 
[T_{\mu\nu}(q_0, p_0, {\mathbf q})]_{q_0=z} \,\, I_z\,,
\label{INTI1}
\end{eqnarray}
where $z$ runs over the poles $\omega_1$, $W_1$, $W_2$,  
$\widetilde \omega_1$, $\widetilde W_1$,
and $\widetilde W_2$ and $[\text{Res} \, I(q_0, \dots) ]_{q_0=z}$ denotes the
 residue of $I(q_0, \dots)$ at the pole $z$. Since the integral 
 $J_{\mu\nu}(p_0,{\mathbf q})$ in a full circle at infinity is zero because  
 the integrand behaves as $q_0^{-4}$ in that limit, closing the Lee-Wick contour 
 upward and including the remaining poles $\omega_2$,  $\widetilde \omega_2$, 
 should yield the same result as Eq.~(\ref{INTI1}). 
In other words, we expect
%................................................Fig3........................................................
%\begin{figure}
%.............................................................................................................
\begin{eqnarray}
J_{\mu\nu}(p_0,{\mathbf q})&=&(2\pi i) \sum_{{\bar z}} T_{\mu\nu}(q_0, p_0, 
{\mathbf q}) \Big|_{q_0={\bar z}} \nonumber \\ &\times& [\text{Res} \, I(q_0, p_0, {\mathbf q})\Big|_{q_0={\bar z}}
\equiv  (2\pi i)\nonumber   \\ &\times& \sum_{\bar z} T_{\mu\nu}(q_0, p_0, {\mathbf q})\Big|_{q_0={\bar z}} \,\, I_{\bar z}\,,
\label{INTI2}
\end{eqnarray}
where ${\bar z}$ runs over the poles $\omega_2$,  $\widetilde \omega_2$. Both $I_z$ and $I_{\bar z}$ 
are functions of $p_0$ and ${\mathbf q}$, which we do not write in the following in order
to keep the notation simple.

Calculating $I_z$ and $I_{\bar z}$ according to the definitions in Eqs.~\eqref{INTI1}
and~\eqref{INTI2}, respectively, we obtain
\begin{eqnarray}
I_{\omega_1}&=& 
\frac{\pi}{ g^2E({\mathbf q}) N_1p_0(\omega_1-\widetilde \omega_2)
(\omega_1-\widetilde W_1)
(\omega_1-\widetilde W_2)}\,, \label{R1}
\nonumber \\
 I_{\widetilde \omega_1}&=& 
\frac{-\pi}{g^2 E({\mathbf q}) N_1p_0(\widetilde \omega_1-\omega_2)
(\widetilde \omega_1-W_1)(
\widetilde \omega_1-W_2)}\,,   \label{R2} 
\nonumber \\
I_{W_1}&=& 
\frac{-\pi}{g^2 E({\mathbf q}) N_1 p_0(W_1-\widetilde \omega_1)
(\widetilde W_1- \omega_2)(\widetilde W_1-W_2)}\,, \label{R3}  
\nonumber \\
I_{\widetilde W_1}&=&
\frac{\pi}{g^2 E({\mathbf q}) N_1 p_0(\widetilde W_1- \omega_1)
(\widetilde W_1- \omega_2)(\widetilde W_1-W_2)}\,,   \label{R4}   
\nonumber \\
 I_{W_2}&=& 
\frac{\pi}{ g^2 E({\mathbf q}) N_2 p_0(W_2-\widetilde \omega_1)
(W_2-\widetilde \omega_2)(W_2-\widetilde W_1)}\,,  \label{R5} 
\nonumber \\
 I_{\widetilde W_2}&=& 
\frac{-\pi}{ g^2 E({\mathbf q}) N_2 p_0(\widetilde W_2-\omega_1)
(\widetilde W_2-\omega_2)(\widetilde W_2-W_1)}\,, \label{R6} 
\nonumber \\
 I_{\omega_2}&=& 
\frac{\pi}{g^2 E({\mathbf q}) N_2 p_0(\omega_2-\widetilde \omega_1)
(\omega_2-\widetilde W_1)(\omega_2-\widetilde
W_2)}\,,  \label{R7}
\nonumber \\
I_{\widetilde \omega_2}&=&
 \frac{-\pi}{g^2 E({\mathbf q}) N_2 p_0(\widetilde \omega_2-\omega_1)(\widetilde 
 \omega_2-W_1)(\widetilde \omega_2-W_2)}\,,  \label{R8} \nonumber \\
\end{eqnarray}
where we have repeatedly used Eqs.~\eqref{ID1} and~\eqref{ID3}.
In the equations above,  the eigenvalues $\omega_1, 
\omega_2, W_1$ and $W_2$ are all functions of $E({\mathbf q})$, 
according to the definitions~\eqref{ROOTS1} and~\eqref{ROOTS2}.  
%................................................................................

In order to compute the imaginary part of ${\cal M}_F$,
we recall that
\be
{\cal M}_F=-\frac{e^4}{p^4}J_1^\nu J_2^\mu \int \frac{d^3q}{(2 \pi)^4}
\sum_z T_{\mu\nu}(q_0,p_0, {\mathbf q})\Big|_{q_0=z} \,\, I_z\,,
\label{AMP3}
\ee
and  we use the relation
\begin{eqnarray}
&&{\rm Im} {\cal M}_F(p_0+i\epsilon)=\frac{1}{2i}\text{Disc}({\cal M}_F)
\nonumber \\ &\equiv& \frac{1}{2i}\left({\cal M}_F(p_0+i\epsilon)
-{\cal M}_F(p_0-i\epsilon)\right)\,.
\label{DEFIP}
\end{eqnarray}
The discontinuity in Eq.~(\ref{AMP3}) arises only from each 
contribution $I_z, I_{\bar z}$. 
Assuming that one of them occurs at $p_0=\alpha$, we focus on the 
relevant part of $I_z$, which we write as
\be
I_z=U(p_0, {\mathbf q}, \dots)\frac{1}{(p_0-\alpha)}\,.
\ee
The  contribution to the discontinuity will be
\begin{eqnarray}
\text{Disc}(I_z)&=&U(p_0=\alpha, {\mathbf q})\frac{1}{2i}  
\nonumber \\ &\times&  \left( \frac{1}{p_0-\alpha+i\epsilon}
 -\frac{1}{p_0-\alpha-i\epsilon} \right)\nonumber \\ &=&\pi\, 
 U(p_0=\alpha, {\mathbf q})\; \delta(p_0-\alpha)\,,
\end{eqnarray}
according to the identity
\begin{equation}
\frac{\epsilon }{x^{2}+\epsilon ^{2}}=\pi \delta (x)\,.
\end{equation}
Next, we list the contributions to the discontinuity arising from each $I_z$ 
and leave to the Appendix the detailed derivation of the results:
\begin{eqnarray}
\text{Disc}(I_{\omega_1})&=&  \frac{i \pi^2}{E^2({\mathbf q})N_1 N_2}
 \delta\left(p_0-(\omega_1-\omega_2)\right)\,,        \label{D1}    
\nonumber \\
\text{Disc}(I_{\widetilde \omega_1})&=& \frac{-i \pi^2}{E^2({\mathbf q})(N_1)^2 } 
\delta(p_0-(W_1-\omega_1))\,,      \label{D2}    
\nonumber  \\
\text{Disc}(I_{W_1})&=&\frac{i \pi^2}{E^2({\mathbf q})(N_1)^2 }
  \delta(p_0-(W_1-\omega_1))\,, \label{D3}    
\nonumber  \\
\text{Disc}(I_{\widetilde W_1})&=& \frac{-i \pi^2}{E^2({\mathbf q}) N_1 N_2 }  
\delta(p_0-(\omega_1-\omega_2))\,,  \label{D4} 
\nonumber  \\
\text{Disc}(I_{W_2})&=& \frac{i \pi^2}{E^2({\mathbf q}) N_1 N_2 }
 \delta(p_0-(\omega_1-\omega_2))\,, \label{D5}    
\nonumber  \\
\text{Disc}(I_{\widetilde W_2})&=& 0\,,  \label{D6}   
\nonumber  \\
\text{Disc}(I_{\omega_2})&=&0\,,  \label{D7}  
\nonumber  \\
\text{Disc}(I_{\widetilde \omega_2})&=& \frac{i \pi^2}{E^2({\mathbf q}) N_1 N_2 }
 \delta(p_0-(\omega_1-\omega_2))\,.  \label{D8} 
\end{eqnarray}
The next step is to calculate the  discontinuities in ${\cal M}$, from the general expression 
\begin{eqnarray}
\text{Disc}({\cal M}_F)&=&-\frac{e^4}{p^4}J_1^\mu(p_1) J_2^\nu(p_2) \int \frac{d^3q}{(2 \pi)^4}
\nonumber \\ &\times& \sum_z T_{\mu\nu}(q_0,p_0, {\mathbf q})\Big|_{q_0=z} \,\, \text{Disc}(I_z)\,.
\label{AMP4}
\end{eqnarray}
Next we concentrate on the basic ingredient
\be
 U_{\mu\nu}({\mathbf q})\Big|_{z}\equiv T_{\mu\nu}(q_0,p_0, {\mathbf q})\Big|_{q_0=z} \times \text{Disc}(I_z)\,.
\label{DEFU}
\ee
where we introduce the further notation
\begin{eqnarray}\label{TMUNUADD}
T_{\mu\nu}(q_0,p_0, {\mathbf q})=\text{Tr}\left( \gamma _{\mu }
(\gamma^0   f(q_0-p_0)+ \gamma^i {  q_i}    +m)\right.  \\  \left. \gamma _{\nu }  
(\gamma^0f(q_0)+ \gamma^iq_i +m) \right)\equiv F_{\mu\nu}(f(q_0-p_0), f(q_0))\,,\nonumber 
\end{eqnarray}
in order to emphasize the relevant variables at this stage. 

According to Eq.~\eqref{AMP4} together with the definitions~\eqref{DEFU} 
and~\eqref{TMUNUADD}, in order to calculate $\text{Disc}({\cal M}_F)$, we need
  the combinations $F_{\mu\nu}(f(q_0-p_0), f(q_0)) \times \text{Disc}(I_z)$. The 
  relevant terms here are the products of $F_{\mu\nu}(f(q_0-p_0), f(q_0))$ times
the delta functions appearing in each $\text{Disc}(I_z)$ arising from Eq.~\eqref{D1}. 
We provide a list of them in the following:
%%%%%%%%%%%%%%%%%%%%%%%%%%%%%%%%%%%%
\begin{widetext}
\begin{eqnarray}
F_{\mu\nu}(f(q_0-p_0), f(q_0))\Big|_{q_0=\omega_1}  \delta(p_0-(\omega_1-\omega_2))&=&
F_{\mu\nu}(-E({\mathbf q}),E({\mathbf q})) 
\delta(p_0-(\omega_1-\omega_2))\,,  \label{FD1}
\nonumber \\
F_{\mu\nu}(f(q_0-p_0), f(q_0))\Big|_{q_0=\widetilde \omega_1}  \delta(p_0-(W_1-\omega_1))&=&
F_{\mu\nu}(E({\mathbf q}),E({\mathbf q})) \delta(p_0-(W_1-\omega_1))\,, \label{FD2}
\nonumber \\
F_{\mu\nu}(f(q_0-p_0), f(q_0))\Big|_{q_0=W_1} \delta(p_0-(W_1-\omega_1))&=&
F_{\mu\nu}(E({\mathbf q}),E({\mathbf q})) \delta(p_0-(W_1-\omega_1))\,,  \label{FD3} 
\nonumber \\
F_{\mu\nu}(f(q_0-p_0), f(q_0))\Big|_{q_0=\widetilde W_1} \delta(p_0-(\omega_1-\omega_2))&=&
F_{\mu\nu}(E({\mathbf q}),-E({\mathbf q})) \delta(p_0-(\omega_1-\omega_2))\,, \label{FD4}
\nonumber \\
F_{\mu\nu}(f(q_0-p_0), f(q_0))\Big|_{q_0=W_2} \delta(p_0-(\omega_1-\omega_2))&=& 
F_{\mu\nu}(E({\mathbf q}),-E({\mathbf q})) \delta(p_0-(\omega_1-\omega_2))\,, \label{FD5} 
\nonumber \\
F_{\mu\nu}(f(q_0-p_0), f(q_0))\Big|_{q_0=\widetilde \omega_2} \delta(p_0-(\omega_1-\omega_2))&=&
F_{\mu\nu}(-E({\mathbf q}),E({\mathbf q})) \delta(p_0-(\omega_1-\omega_2))\,.  \label{FD6} 
\end{eqnarray}
\end{widetext}
In proving the above results, we have extensively used the relations (\ref{RELADD}).
In this way, the contributions to the discontinuity of ${\cal M}$ are
\begin{eqnarray}
U_{\mu\nu}({\mathbf q})\Big|_{\omega_1}&=&\frac{i \pi^2}{E^2({\mathbf q})N_1 N_2}
F_{\mu\nu}\left(-E({\mathbf q}),E({\mathbf q}) \right)\nonumber \\ &\times& \delta(p_0-(\omega_1-\omega_2))\,,  \label{U11}
\end{eqnarray}
\begin{eqnarray}
U_{\mu\nu}({\mathbf q})\Big|_{\widetilde \omega_1}&=&-\frac{i \pi^2}{E^2({\mathbf q})(N_1)^2 }
F_{\mu\nu}\left(E({\mathbf q}),E({\mathbf q})\right) \nonumber  \\ &\times&
\delta(p_0-(W_1-\omega_1)) \label{U21}\,, 
\end{eqnarray}
\begin{eqnarray}
U_{\mu\nu}({\mathbf q})\Big|_{W_1}&=&\frac{i \pi^2}{E^2({\mathbf q})(N_1)^2 }
F_{\mu\nu}\left(E({\mathbf q}),E({\mathbf q})\right) \nonumber 
 \\ &\times& \delta(p_0-(W_1-\omega_1)),  \label{U31}\,, 
\end{eqnarray}
\begin{eqnarray}
U_{\mu\nu}({\mathbf q})\Big|_{\widetilde W_1}&=&-\frac{i \pi^2}{E^2({\mathbf q}) N_1 N_2 } 
 F_{\mu\nu}\left(E({\mathbf q}),-E({\mathbf q})  \right) \nonumber   \\ &\times&
\delta(p_0-(\omega_1-\omega_2))  \label{U41}\,,
\end{eqnarray}
\begin{eqnarray}
U_{\mu\nu}({\mathbf q})\Big|_{W_2}&=& \frac{i \pi^2}{E^2({\mathbf q}) N_1 N_2 }
F_{\mu\nu}\left(E({\mathbf q}), -E({\mathbf q})  \right)\nonumber    
\\ &\times& \delta(p_0-(\omega_1-\omega_2))\,, \label{U51}
\end{eqnarray}
\begin{eqnarray}
U_{\mu\nu}({\mathbf q})\Big|_{\widetilde W_2}=0\,,\label{U61}
\end{eqnarray}
\begin{eqnarray}
U_{\mu\nu}({\mathbf q})\Big|_{\omega_2}=0\,, \label{U71} 
\end{eqnarray}
\begin{eqnarray}
U_{\mu\nu}({\mathbf q})\Big|_{\widetilde \omega_2}&=&\frac{i \pi^2}{E^2({\mathbf q}) N_1 N_2 }
F_{\mu\nu}\left(-E({\mathbf q}), E({\mathbf q})\right)  \nonumber \\ &\times&  \delta(p_0-(\omega_1-\omega_2))\,. \label{U81}  
\end{eqnarray}
Let us observe that unexpected cancellations occur:
\begin{eqnarray}
U_{\mu\nu}({\mathbf q})\Big|_{\widetilde \omega_1}+U_{\mu\nu}({\mathbf q})\Big|_{W_1}&=&0,\nonumber  \\ 
U_{\mu\nu}({\mathbf q})\Big|_{\widetilde W_1}+ U_{\mu\nu}({\mathbf q})\Big|_{W_2} &=& 0.
\label{CANCEL}
\end{eqnarray}
Now we are in position to calculate the final result for $\text{Disc}({\cal M}_F)$, 
in agreement with the Lee-Wick contour $C_{LW}$ previously 
chosen. When we evaluate the integral over $q_0$ in $J_{\mu\nu}(p_0, {\mathbf q})$, 
Eq.~{\eqref{DEFJMUNU}, closing $C_{LW}$ from below, we get the contributions from the poles  
$ z: \omega_1, W_1$ and $W_2$,
plus those obtained by the displacement in $p_0$, according to Eq.~\eqref{INTI1}. This 
means that only the corresponding contributions from $U_{\mu\nu}({\mathbf q})\Big|_z$ 
add up in Eq.~{\eqref{AMP4}. Considering further that  $U_{\mu\nu}({\mathbf q})\Big|_{\widetilde W_2}=0$
together with the cancellations in Eq.~\eqref{CANCEL}, the final  contribution to $\text{Disc}({\cal M}_F)$
comes only from  $U_{\mu\nu}({\mathbf q})\Big|_{\omega_1}$, with the result 
\begin{eqnarray}
\frac{1}{i}\text{Disc}({\cal M}_F)&=&\frac{e^4}{p^4}J_1^\nu J_2^\mu \int \frac{d^3q}{(2 \pi)^2}
\frac{1}{2E({\mathbf q})N_1 2E({\mathbf q}) N_2} \nonumber \\ &&
\text{Tr}\left(\gamma_{\mu }
(\gamma^0 {E({\mathbf q})}- \gamma^i {q_i} - m) \right.\nonumber  \\ && \left. \gamma_{\nu }
(\gamma^0{E({\mathbf q})}+\gamma^iq_i +m)\right) \nonumber \\ 
&& \times \delta(p_0-(\omega_1-\omega_2))
\label{AMP6}
\end{eqnarray}
in the center-of-mass frame. If we were to close the Lee-Wick contour from 
above in  Eq.~\eqref{DEFJMUNU}, the relation $\eqref{INTI2}$ tells us that nothing 
but the poles ${\bar z} : \, \omega_2, \, \widetilde \omega_2 $ need to be included. Then, the 
corresponding contributions to $U_{\mu\nu}({\mathbf q})\Big|_{\bar z}$ arise only 
from $U_{\mu\nu}({\mathbf q})\Big|_{\widetilde \omega_2 }$
in Eq.~\eqref{U81}, which is equal to  $U_{\mu\nu}({\mathbf q})\Big|_{\omega_1}$. In this way, 
we have explicitly shown that the result in Eq.~\eqref{AMP6} is independent of the way in which 
we calculate the $q_0$ integral in Eq.~\eqref{DEFJMUNU}. As mentioned 
previously, this is a consequence  of the fact that such an integral is zero in a circle at infinity.
%....................................................................
\section{Verification of the Optical Theorem}
\label{VEROPT}
We have already calculated the left-hand side of Eq.~\eqref{OPTTHEO} in the 
evaluation of  the optical theorem. Now we deal with the contribution of the final  
states required in the right-hand side of this equation.
To the order considered, we have only two-particle final states. In this way, we 
start by calculating the amplitude ${\mathcal M}_I$ for the process
$e^{-}(p_{1},s)+e^{+}(p_{2},r)\rightarrow
e^{-}(k_{1},\bar{s})+e^{+}(k_{2},\bar{r})$.
As already stated, we apply the Lee-Wick prescription in such a way that we only   
consider the asymptotic states corresponding to those with a
 positive metric, corresponding to the frequencies 
  $\omega_1$ and $\omega_2$. We have also defined our effective model 
to have $ E({\mathbf p})< 1/4g$, in order to deal only with real frequencies
at which the Hamiltonian is stable and hermitian.
We obtain 
\be
\mathcal{M}_I=-\frac{ie}{p^{2}}J_{1}^{\mu}\left[ \bar{v}^{
\bar{r}}(k_{2})(-ie\gamma _{\mu })u^{\bar{s}}(k_{1})\right] \frac{1}{\sqrt{
N_{1}(\mathbf{k}_{1})N_{2}(\mathbf{k}_{2})}},
\label{RHS1}
\ee
where we have introduced the current defined in Eq.~(\ref{CURRENTS}). This yields 
\begin{eqnarray}
|\mathcal{M}_I|^{2}&=&\frac{e^{2}}{p^{4}}\frac{1}{N_{1}(\mathbf{k}
_{1})N_{2}(\mathbf{k}_{2})}J_{1}^{\mu }(p_{1},p_{2})   \left[ J_{1}^{\alpha
}(p_{1},p_{2})\right] ^{\ast }\nonumber  \\ && \left[ \bar{v}^{\bar{r}}(k_{2})(e\gamma _{\mu })u^{\bar{s}}(k_{1})
\right] \bar{u}^{\bar{s}}(k_{1})(e\gamma _{\alpha })v^{\bar{r}}(k_{2})\,.\nonumber \\
\label{RHS2}
\end{eqnarray}
Recalling the right-hand side of Eq.~(\ref{OPTTHEO}),which we denote by  $W$, 
\begin{eqnarray}
W&=&\sum_{\bar{r},\;\bar{s}}\int \frac{d^{3}k_{1}}{(2\pi )^{3}}\frac{1}{2E(
\mathbf{k}_{1})}\frac{d^{3}k_{2}}{(2\pi )^{3}}\frac{1}{2E(\mathbf{k}_{2})}
(2\pi )^{4} \nonumber \\ &\times&  \delta ^{4}(p=k_{1}+k_{2})|\mathcal{M}_I|^{2},
\label{RHS3}
\end{eqnarray}
we perform the sum over the spin components ${\bar s}, {\bar r}$ 
with the result
\begin{eqnarray}
\sum_{\bar{r},\;\bar{s}}|\mathcal{M}_I|^{2}&=&\frac{1}{p^{4}}\frac{1}{
N_{1}(\mathbf{k}_{1})N_{2}(\mathbf{k}_{2})}J_{1}^{\mu }(p_{1},p_{2})\left[
J_{1}^{\alpha }(p_{1},p_{2})\right] ^{\ast }\nonumber  \\ && \text{Tr}\left[ 
\gamma _{\mu }\left( \gamma k_{1}+m\right) \gamma _{\alpha
}\left( \gamma k_{2}-m\right) \right]. 
\label{RHS4}
\end{eqnarray}
In the center-of-mass frame, we have
%\begin{widetext}
\begin{eqnarray}
W&=&\frac{e^{4}}{p^{4}}J_{1}^{\mu }(p_{1},p_{2})\left[ J_{1}^{\alpha
}(p_{1},p_{2})\right] ^{\ast } \nonumber \\  &\times&\int \frac{d^{3}k_{1}d^{3}k_{2}}{(2\pi )^{2}2E(\mathbf{k}_{1})N_{1}(\mathbf{k}_{1})2E(\mathbf{k}_{2})N_{2}(\mathbf{k}_{2})} \nonumber
\\  &\times& \delta ^{3}(\mathbf{k}_{1}+\mathbf{k}_{2})  
\delta \left(p_{0}-(\omega _{1}-\omega
_{2})\right)  \nonumber   \\ &\times&\text{Tr}\left[ \gamma _{\mu }    \left( \gamma ^{0}E(\mathbf{k}_{1}) +\gamma
^{i}k_{1i}+m\right)  \gamma _{\alpha }\left( \gamma ^{0}E(\mathbf{k}%
_{2})  \right. \right. \nonumber  \\  &+& \left. \left. \gamma ^{i}k_{2i}-m\right) \right]\,.
\label{RHS5}
\end{eqnarray}
In the last step, we relabel ${\mathbf k}_1 \rightarrow
 {\mathbf q} $,  and we integrate over $d^{3}k_{2}$. In this way,
\begin{eqnarray}
W &=&\frac{e^{4}}{p^{4}}J_{1}^{\mu }(p_{1},p_{2})\left[ J_{1}^{\alpha
}(p_{1},p_{2})\right] ^{\ast }\nonumber \\  &\times&  \int \frac{d^{3}q }{(2\pi )^{2}2E(\mathbf{q}
)N_{1}(\mathbf{q})2E(\mathbf{q})N_{2}(\mathbf{q})}   \nonumber \\  &\times&  \delta \left(p_{0}-(\omega _{1}(
\mathbf{q})-\omega _{2}(\mathbf{q}))\right) \nonumber \\ &\times&
\text{Tr}\left[ \gamma _{\mu }\left( \gamma ^{0}E(\mathbf{q})+\gamma
^{i}q_{i}+m\right) \gamma _{\alpha }\left( \gamma ^{0}E(\mathbf{q}) \right. \right. \nonumber  \\  &-& \left. \left.\gamma
^{i}q_{i}-m\right) \right]. 
\label{RHS6}
\end{eqnarray} 
%\end{widetext}
The cyclic property of the trace together with the relation $[J_1^\alpha]^\ast=
J_2^\alpha$ from Eq.~\eqref{CURRENTS} show that $W=2 {\rm Im}[{\cal M}_F]
=-i \text{Disc}({\cal M}_F)$, where the last expression  is given in Eq.~\eqref{AMP6}, thus 
verifying the optical theorem.
%-------------------------------------------------------------
\section{Conclusions and outlook}
\label{CONCL}
The effective approach to quantum field theory 
provides a powerful tool to search for new physics beyond the standard model.
In particular, the search for quantum gravity effects at low energies in the form of 
Lorentz symmetry violations has become an active research area from both
the
phenomenological
and experimental points of view.
The majority
of these searches have been realized by coupling constant tensors, yielding
Lorentz invariance violation, with derivative operators of renormalizable mass dimension.
In this way one guarantees from the beginning some crucial
requirements about stability and unitarity in the effective quantum field theory. 

However, the study of Lorentz symmetry violation incorporating higher-order 
derivative operators has attracted interest in the last few years.
There are good reasons for this: (i) Bounds arising from
higher-order derivative operators have been less explored experimentally
as compared to those arising from renormalizable models, and (ii)
Due to the increase of the number of degrees of freedom in models with
higher-order derivative operators, they have the potentiality
to capture higher-energy degrees of freedom associated with
new physics. 
Also, it is well known that the introduction of
higher-order derivative operators has the advantage of
 smoothing ultraviolet divergences.
The nonminimal SME and the Myers-Pospelov model
provide the framework to detect possible effects of
higher-order Lorentz invariance violation. 
One of the main goals of this work is to emphasize that when 
dealing with models described by operators with mass dimension greater than four,
it is also necessary to consider the consequences upon unitarity, as shown in the past works 
of Lee and Wick.

The Lee-Wick extension of quantum electrodynamics is
a modified quantum field theory with an indefinite metric. 
One of the goals in this construction has been to prove the preservation of 
 unitarity by applying what is now called the Lee-Wick prescription~\cite{LW}.
 This prescription consists in removing the 
 negative-metric
states from the asymptotic Hilbert space and in redefining the contour integration
in Feynman diagrams 
to preserve the perturbative unitarity of the $S$ matrix. The prescription has 
to be defined order by order in the perturbative series expansion.

In this work, we have followed the Lee-Wick prescription in a model
where Lorentz violation is explicitly broken with a preferred four-vector coupled to 
a higher-dimension derivative operator. In particular, we have focused on the 
dimension-five operators
of an effective Myers-Pospelov model, which is quantized in a box with boundary conditions and specific prescriptions which preclude the appearance of imaginary frequencies, imposing the restriction $E({\mathbf p}) < 1/4g$ for $0 < |{\mathbf p}| < \infty$.
Our goal has been to test the Lee-Wick prescription in order to verify perturbative unitarity
in the form of the optical theorem. 
We have found that
 unitarity is preserved at the one-loop level in the annihilation channel of the 
 scattering process $e^++e^-$.
 The computation has taken into 
 consideration the discontinuities of the Feynman diagram on the left-hand side of Fig.~(\ref{Fig2}), for energies $E({\mathbf p})$ below the critical value $1/4g$, together with a definition of the adequate Lee-Wick contour $C_{LW}$.
 
A further step in generalizing this result will be to probe the 
optical theorem in each of the remaining
one-loop
Feynman diagrams that appear in the model,
for the same scattering processes.
%---------------------------------
\section*{Acknowledgments}
C.M.R. acknowledges partial support 
by the research project Fondecyt Regular No.\ 1140781-Chile
and the project group of \emph{F\'{\i}sica de Altas Energ\'{\i}as}
of the Universidad del B{\'i}o-B{\'i}o, and wants to thank Luis Urrutia for his
hospitality at the Universidad Nacional Aut\'onoma
de M\'exico. L.F.U.  has been partially supported by Project No. IN104815 from 
Direcci\'on General Asuntos del
Personal Acad\'emico (Universidad Nacional Aut\'onoma
de M\'exico) and by Project CONACyT (M\'exico) No. 237503.
%.................................Appendix........................................
\appendix
\section{THE CONTRIBUTIONS TO THE DISCONTINUITY OF $I(p_0, {\mathbf q})$}
%...................................................................................
For each term $I_z$ or $I_{\bar z}$, we indicate the possible 
contributions to the discontinuity [the choices of 
$p_0= Y(|{\mathbf q}|) $ which make each denominator zero]
and analyze which such conditions can be 
fulfilled. On the one hand, we have
\begin{eqnarray}
p_0(|{\mathbf p}|)&=&\omega_1(|{\mathbf p}|)+
|\omega_2(|{\mathbf p}|)| \nonumber \\ &=&\frac{\sqrt{1+4gE({\mathbf p})}-
\sqrt{1-4gE({\mathbf p})}}{2g}  \nonumber  \\ &=&\frac{N_2(|{\mathbf p}|)-N_1(|{\mathbf p}|)}{2g}
\label{P0}
\end{eqnarray} 
in the center-of-mass frame, where ${\mathbf p}_{e^+}=-{\mathbf p}_{e^-}= {\mathbf p}$.
On the other hand, the condition for a discontinuity to occur
 is that the equation $p_0(|{\mathbf p}|)=Y(|{\mathbf q}|)$ 
 have a solution for $|{\mathbf q}|$. The function $Y(|{\mathbf q}|)$ 
 will depend on the various combinations of the eigenvalues 
 $\omega_1, \omega_2, W_1$ and $W_2$ defined in 
 Eqs.~\eqref{ROOTS1} and~\eqref{ROOTS2}. We have 
 to consider only the positive contributions to $p_0$, which we 
 discuss below, in order to further evaluate 
 the discontinuities arising  from Eq.~\eqref{R1}. 
%...................................................................................
\subsection{Identification of the contributions to the discontinuity}
\label{IDDISC}
The following cases arise:

{\bf Case 1:}
\be
  p_0=\frac{N_2({\mathbf q})-N_1({\mathbf q})}{2g}\,,
\label{CASE1}
\ee
which is directly solved by 
choosing ${\mathbf q}={\mathbf p}$ according to Eq.~\eqref{P0}\,.

{\bf Case 2:}
\be
p_0=\frac{N_2({\mathbf q})+N_1({\mathbf q})}{2g}
=\frac{\sqrt{1+4gE({\mathbf q})}+\sqrt{1-4gE({\mathbf q})}}{2g}\,.
\label{CASE2}
\ee
Substituting Eq.~\eqref{P0} and taking 
the square of the resulting equation, we obtain
\be
\sqrt{1-16g^2E^2({\mathbf p})}=-\sqrt{1-16g^2E^2({\mathbf q})}\,,
\ee 
which produces  a sign inconsistency, leading to no solution in this case.

{\bf Case 3:}
\be
p_0=\frac{N_2({\mathbf q})}{g}=\frac{\sqrt{1+4gE({\mathbf q})}}{g}\,.
\label{CASE3}
\ee
Replacing $p_0$ as before and taking the square of the resulting equation yields
\be
-\sqrt{1-16 g^2 E^2({\mathbf p})}=1+8 g E^2({\mathbf q})\,.
\ee
The left-hand side of the above equation is negative, while the right-hand 
side is positive, leading again to no solution for $|{\mathbf q}|$.

{\bf Case 4: }
\be
p_0=\frac{N_1({\mathbf q})}{g}=\frac{\sqrt{1-4gE({\mathbf q})}}{g}\,.
\label{CASE4}
\ee
Replacing $p_0$ as before and taking the square of the resulting equation yields
\be
-\sqrt{1-16 g^2 E^2({\mathbf p})}=1-8 g E^2({\mathbf q})\,.
\label{CASE44}
\ee
Since $4 g E^2({\mathbf q})< 1$, we still can have a solution in the region 
\be
1 < 8 g E^2({\mathbf q}) < 2.
\ee 
In this case the right-hand side of Eq.~\eqref{CASE44} 
is negative. Solving for the resulting equation, we get
\be 
E^2({\mathbf q})=\frac{1+\sqrt{1-16 g^2 E^2({\mathbf p})}}{8g}\,.
\label{CASE444}
\ee 
In fact, Eq.~\eqref{CASE444} is satisfied for the whole 
range of values of $E({\mathbf p})$, while for 
$E({\mathbf p=0})$, we have $ 8gE^2({\mathbf q})={
1+\sqrt{1-16 g^2 m^2}} < 2$, while for For $E({\mathbf p})
=E_{\rm max}=1/4g $ we obtain $8gE^2(q)=1$. Thus,
this case will contribute to the discontinuity.
%.............................................................................................
\subsection{The particular cases}
To compute ${\rm Disc}(I_{\omega_1})$ from the first
Eq.~\eqref{R1}, we have the possible choices for $p_0$
\begin{eqnarray}
p_0&=&\omega_1-\omega_2=\frac{N_2-N_1}{2g}, \label{REL1} \\ 
 p_0&=&\omega_1-W_1=-\frac{N_1}{g} <0\,, 
 \end{eqnarray}
 and
 \begin{eqnarray}    
p_0=\omega_1-W_2=-\frac{N_1+N_2}{2g} < 0\,,
\end{eqnarray}
where we have used Eqs.~\eqref{ROOTS1} and~\eqref{ROOTS2}.  
Since $ N_2 > N_1 > 0$, the only contribution
 arises from the first case in Eq.~\eqref{REL1}, which yields
\begin{eqnarray}
\text{Disc}(I_{\omega_1})&=&i (2\pi )^2\frac{\delta(-\omega_1+\omega_2+p_0)}{g^2 2E({\mathbf q})N_1}
 \\ &\times&
  \frac{1}{p_0
(\omega_1-W_1-p_0)(\omega_1-W_2-p_0) }\,.\nonumber
\end{eqnarray}
The delta function allows us to rewrite the denominator as
\begin{eqnarray}
  \frac{-1}{g^2(  \omega_2-\omega_1)
(\omega_2-W_1)(\omega_2-W_2)}\,.
\end{eqnarray}
Using  the relation~\eqref{ID3}
we finally  arrive at
\begin{equation}
\text{Disc}(I_{\omega_1})= \frac{(i \pi^2)}{E^2({\mathbf q}) N_1 N_2 } \delta(p_0-(\omega_1-\omega_2))\,. 
\end{equation}

The next calculation for $Disc(I_{\omega_1+p_0})$ follows 
closely the previous case, so we mention only the relevant points.
From Eq.~\eqref{R2}, we read the following possible values for $p_0$:
\begin{eqnarray}
p_0&=&-(\omega_1-\omega_2)=-\frac{N_2-N_1}{g} < 0\,, 
\nonumber \\ p_0&=&-(\omega_1-W_1)=\frac{N_1}{g}\,, \nonumber \\ 
p_0&=&-(\omega_1-W_2)=\frac{N_1+N_2}{2g}\,,
\label{REL2}
\end{eqnarray}
According to Sec.~\ref{IDDISC},
only the second case in Eq.~\eqref{REL2} survives, yielding
\be
\text{Disc}(I_{\omega_1+p_0})=-\frac{i\pi^2}{E^2({\mathbf q}) (N_1)^2}\delta(p_0-(W_1-\omega_1))\,.
\ee

For $\text{Disc}(I_{W_1})$,
the possibilities for $p_0$ are
\begin{eqnarray}
p_0&=&W_1-\omega_1=\frac{N_1}{g}\,, \nonumber \\ p_0&=&W_1-\omega_2=\frac{N_1+N_2}{2g}\,, 
\nonumber \\ p_0&=&W_1-W_2=\frac{N_1-N_2}{2g} < 0\,.
\label{REL3}
\end{eqnarray}
From Sec.~\ref{IDDISC}, we conclude that the only
 contribution arises from the first case in Eq.~\eqref{REL3}, which produces
\be
\text{Disc}(I_{W_1})=\frac{i \pi^2}{E^2({\mathbf q}) (N_1)^2} \delta(p_0-(W_1-\omega_1))\,.
\ee
Now, we look at $\text{Disc}(I_{\widetilde W_1})$. 
From Eq.~\eqref{R4}, we have the following possible values for $p_0$:
\begin{eqnarray}
p_0&=&-(W_1-\omega_1)=-\frac{N_1}{g} < 0\,,\nonumber  \\p_0&=&-(W_1-\omega_2)=-\frac{N_1+N_2}{2g} < 0, 
\nonumber \\p_0&=&-(W_1-W_2)=\frac{N_2-N_1}{2g}\,.
\label{REL4} 
\end{eqnarray}
From  Sec.~\ref{IDDISC}, we conclude that the only contribution arises from the third term
 in Eq.~\eqref{REL4}. We are left with
\begin{eqnarray}
\text{Disc}(I_{\widetilde W_1})= -\frac{i\pi^2}{E^2({\mathbf q}) N_1 N_2} \delta(p_0-(W_2-W_1))\,. 
\end{eqnarray}
From Eq.~\eqref{REL_FREQ}, we get $W_2-W_1=\omega_1-\omega_2$ so that we can write
\begin{eqnarray}
\text{Disc}(I_{\widetilde W_1})=-\frac{i\pi^2}{E^2({\mathbf q}) N_1 N_2} \delta(p_0-(\omega_1-\omega_2))\,. 
\end{eqnarray}

For $\text{Disc}(I_{W_2})$ the choices for $p_0$ are
\begin{eqnarray}
p_0&=&W_2-\omega_1=\frac{N_2+N_1}{2g},\nonumber \\ p_0&=&W_2-\omega_2= \frac{N_2}{g}\,, \nonumber
\\ 
p_0&=&W_2-W_1= \frac{N_2-N_1}{2g}\,.
\label{REL5}
\end{eqnarray}
The discontinuity arises only from the third contribution of the above equation, yielding
\begin{eqnarray}
\text{Disc}(I_{W_2})=\frac{i\pi^2}{E^2({\mathbf q})N_2 N_1} \delta(p_0-(\omega_1-\omega_2))\,. 
\end{eqnarray}

For $\text{Disc}(I_{\widetilde W_2})$, we have
\begin{eqnarray}
p_0&=&-(W_2-\omega_1)=-\frac{N_2+N_1}{2g} < 0\,, \nonumber 
\\ p_0&=&-(W_2-\omega_2)= -\frac{N_2}{g}\,, \nonumber \\ 
p_0&=&-(W_2-W_1)= -\frac{N_2-N_1}{2g} < 0\,,
\label{REL6}
\end{eqnarray}
in such a way that $\text{Disc}(I_{\widetilde W_2})=0$.

For $\text{Disc}(I_{\omega_2})$, we have
\begin{eqnarray}
p_0&=&\omega_2-\omega_1=\frac{(N_1-N_2)}{2g} < 0\,, \nonumber\\ p_0
&=&\omega_2-W_1=-\frac{(N_1+N_2)}{2g} < 0\,, \nonumber
\\ p_0&=&\omega_2-W_2=-\frac{N_2}{g} < 0.
\label{REL7}  
\end{eqnarray}
Since all the contributions are negative we conclude that $\text{Disc}(I_{\omega_2})$=0.

For $\text{Disc}(I_{\widetilde \omega_2})$, we have
\begin{eqnarray}
p_0&=&-(\omega_2-\omega_1)=\frac{(N_2-N_1)}{2g}\,, \nonumber\\ p_0
&=&-(\omega_2-W_1)=\frac{(N_1+N_2)}{2g} \,, \nonumber \\ p_0&=&-(\omega_2-W_2)=\frac{N_2}{g} < 0\,.  
\label{REL8}
\end{eqnarray}
Only the first term in the previous equations contribute, yielding
\be
\text{Disc}(I_{\widetilde \omega_2})=\frac{i \pi^2}{E^2({\mathbf q}) N_1 N_2} \delta(p_0-(\omega_1-\omega_2))\,.
\ee 
%%%%%%%%%%%%%%%%%%%%%%%%%%%%%%%%%%%%%%%%%%%

\end{document}